\documentclass[structabstract]{aa}  

\usepackage{graphicx}
\usepackage[varg]{txfonts}
\usepackage{amsmath}
\usepackage{longtable}
\usepackage{longtable,lscape}
\usepackage{lscape}
\usepackage{epsfig}
\usepackage{color}

\newcommand\be{\begin{equation}}
\newcommand\en{\end{equation}}

\usepackage{txfonts}
\usepackage{hyperref}
\begin{document}

\title{The 2014-2017 outburst of the young star ASASSN-13db:}

\subtitle{A time-resolved picture of a very-low-mass star between EXors and FUors}

\author{A. Sicilia-Aguilar\inst{1,2}, A. Oprandi\inst{3,2}, D. Froebrich\inst{4},  M. Fang\inst{5}, J.~L. Prieto\inst{6,7}, 
K. Stanek\inst{8,9},  A. Scholz\inst{2}, C.S. Kochanek\inst{8,9},  Th. Henning\inst{10}, R. Gredel\inst{10},  
T.W.-~S. Holoien\inst{7,8}, M. Rabus\inst{11, 10}, B.~J. Shappee\inst{12}\thanks{Hubble, Carnegie-Princeton Fellow}, S. J. Billington\inst{4}, J. Campbell-White\inst{4}, and T.~J. Zegmott\inst{4}}

\institute{\inst{1} SUPA, School of Science and Engineering, University of Dundee, Nethergate, Dundee DD1 4HN, UK \\
        \email{a.siciliaaguilar@dundee.ac.uk}\\
\inst{2} SUPA, School of Physics and Astronomy, University of St Andrews, North Haugh, St Andrews KY16 9SS, UK\\
\inst{3} School of Physics and Astronomy, University of Edinburgh, Peter Guthrie Tait Road, Edinburgh EH9 3FD\\
\inst{4} Centre for Astrophysics \& Planetary Science, School of Physical Sciences, University of Kent, Canterbury CT2 7NH, UK\\
\inst{5} Department of Astronomy, University of Arizona, 933 North Cherry Avenue, Tucson, AZ 85721, USA\\
\inst{6} N\'{u}cleo de Astronom\'{i}a de la Facultad de Ingenier\'{i}a y Ciencias, Universidad Diego Portales, Av. Ej\'{e}rcito 441, Santiago, Chile\\
\inst{7} Millennium Institute of Astrophysics, Santiago, Chile\\
\inst{8} Department of Astronomy, The Ohio State University, 140 West 18th Avenue, Columbus, OH 43210, USA\\
\inst{9} Center for Cosmology and AstroParticle Physics (CCAPP), The Ohio State University, 191 W. Woodruff Ave.,
Columbus, OH 43210, USA\\
\inst{9} Department of Astronomy, The Ohio State University, 4055 McPherson Lab, 140 West 18th Avenue, Columbus, OH 43210, USA\\
\inst{10} Max-Planck-Institut f\"{u}r Astronomie, K\"{o}nigstuhl 17, 69117 Heidelberg, Germany\\
\inst{11} Instituto de Astrof\'isica, Facultad de F\'isica, Pontificia Universidad Cat\'olica de Chile, Av.\ Vicu\~na Mackenna 4860, 
7820436 Macul, Santiago, Chile\\
\inst{12} The Observatories of the Carnegie Institution for Science, 813 Santa Barbara St., Pasadena, CA 91101, USA\\
}

\date{Submitted May 29, 2017. Accepted August 4, 2017}

 
  \abstract
   {Accretion outbursts are key elements in star formation. ASASSN-13db is a M5-type star
with a protoplanetary disk, the lowest-mass star known to experience accretion outbursts. Since 
its discovery in 2013, it has experienced two outbursts, the second of which started in November 2014
and lasted until February 2017.}
   {We explore the photometric and spectroscopic behavior of ASASSN-13db during the 2014-2017 outburst.}
   {We use high- and low-resolution spectroscopy and
time-resolved photometry from the ASAS-SN survey, the LCOGT and the Beacon Observatory to 
study the lightcurve of ASASSN-13db and the dynamical and physical
properties of the accretion flow.}
{The 2014-2017 outburst lasted for nearly 800 days. A 4.15d period in the light curve likely
corresponds to rotational modulation of a star with hot spot(s).
The spectra show multiple emission lines with variable inverse P-Cygni profiles and a highly 
variable blue-shifted absorption below the continuum.
Line ratios from metallic emission lines (Fe I/Fe II, Ti I/Ti II) suggest temperatures of $\sim$5800-6000 K 
in the accretion flow.}
   {Photometrically and spectroscopically, the 2014-2017 event displays an intermediate behavior between EXors 
and FUors. The accretion rate 
(\.{M}=1-3$\times$10$^{-7}$M$_\odot$/yr),
about two orders of magnitude higher than the accretion rate in quiescence,
is not significantly different from the accretion rate observed in 2013. 
The absorption features in the spectra suggest that the system 
is viewed at a high angle and drives a powerful, non-axisymmetric wind, maybe
related to magnetic reconnection.
The properties of ASASSN-13db suggest that temperatures lower than those for solar-type stars are needed for modeling
accretion in very-low-mass systems. 
Finally, the rotational modulation during the outburst
reveals that accretion-related structures settle after the beginning of the outburst and 
can be relatively stable and long-lived. 
Our work also demonstrates the power of time-resolved photometry and spectroscopy to explore the
properties of variable and outbursting stars.}

\keywords{stars: pre-main sequence, stars: variability, stars: individual (ASASSN-13db, SDSS J051011.01-032826.2), protoplanetary disks, accretion, techniques: spectroscopic, stars:low-mass}

\authorrunning{Sicilia-Aguilar et al.}

\titlerunning{ASASSN-13db 2014-2017 outburst}

\maketitle

%

\section{Introduction \label{intro}}

Variability is one of the defining characteristics of young
T Tauri stars \citep[TTS;][]{joy45}. Together with rotational modulation due to stellar spots and
extinction by circumstellar material, changes in the accretion rate are one of the reasons
for their variability \citep{herbst94}. Although most TTS appear to undergo only
mild accretion variations on timescales of days to years \citep[e.g.,][]{sicilia10,costigan14},
the accretion rates of some TTS can change by several orders of magnitude
on timescales of weeks to decades. Such eruptive variables are classified as FUors and EXors,
named after their respective prototypes FU Orionis \citep{herbig77,herbig89,hartmann96}
and EX Lupi \citep{herbig01,herbig08}. Accretion outbursts play an important role in the formation
of stars, and may be the key to solving the protostellar luminosity problem \citep{kh95,dunham12} and the formation
of cometary material in the Solar System \citep{abraham09}.
The distinction between the two classes lies in the magnitude of the outburst, the
increase of accretion, the shape of the light curve, and the spectral features observed during outburst.
The characteristics of individual objects do not always fully coincide with one
of the classes \citep{herczeg16}, and some authors have suggested that the two 
classes (or at least, a subset of them) are part of a
continuous spectrum of outbursting stars \citep{contreras14,contreras17}.

\begin{figure*}
\centering
\includegraphics[width=16cm]{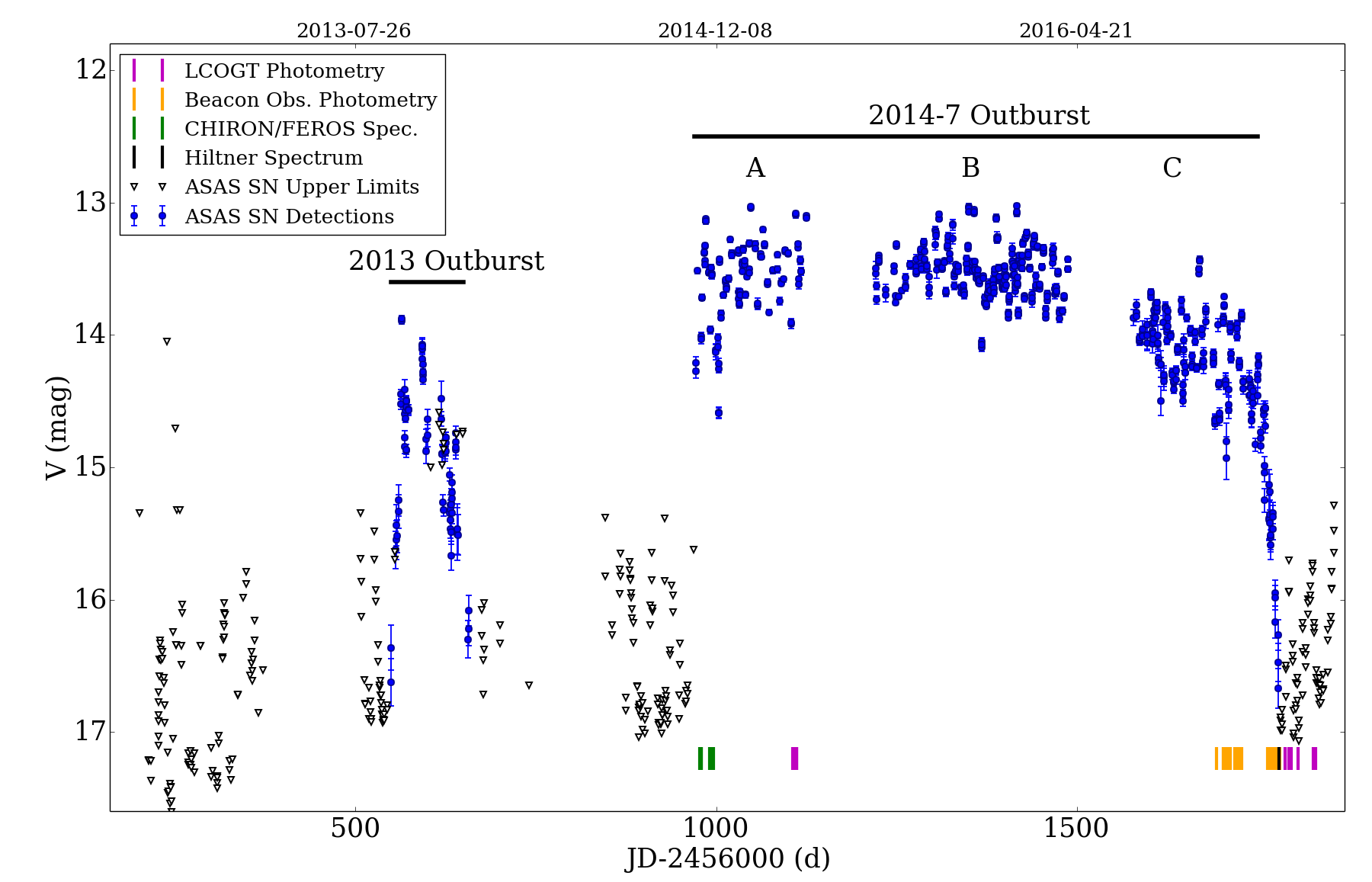}
\caption{V-band ASASSN-13db light curve from ASAS-SN. Detections and non-detections are shown as dots (with errors) and inverted
triangles, respectively. The dates at which we acquired the rest of the observations and the periods of the two known 
outbursts are also marked. }
\label{ASAScurve-fig}%
\end{figure*}

The low-mass star SDSS J05101100-0328262, also known as SDSSJ0510 and ASASSN-13db \citep{holoien14}, 
is a variable star that was identified 
by the All Sky Automated Survey for SuperNovae (ASAS-SN\footnote{http://www.astronomy.ohio-state.edu/$\sim$assassin/index.shtml})
after a four-magnitude brightness increase 
in September 2013 \citep{shappee14,holoien14}. 
Spectroscopic observations during the 2013 outburst 
revealed a rich emission line spectrum, leading
to its classification as an EXor variable \citep{holoien14}.
The spectrum contained hundreds of metallic emission lines,  so that
the object was dubbed "the EX Lupi twin", entering the category as one of the most 
impressive EXor variables considering its photometric variability and spectral features \citep{holoien14}.
Being a very red star with substantial IR emission
consistent with a protoplanetary disk \citep{holoien14}, 
it is likely a member of the young star-forming regions within the
L1615/L1616 Orion cometary clouds. The regions are part of the open cluster NGC 1981 to the north of the Orion 
Nebula Cluster (ONC).  ASASSN-13db would have an approximate age of 1-3 Myr
based on other stars in the same region \citep{gandolfi08}.
Further observations during quiescence in January 2014 confirmed that ASASSN-13db is a young, accreting TTS
M5-type star, which also makes it the lowest-mass EXor identified to date \citep{holoien14}.
After a brief period of quiescence during 2014, ASASSN-13db went into outburst again in November 2014
\citep[ASAS SN CV Patrol\footnote{http://cv.asassn.astronomy.ohio-state.edu/};][]{davis15}.

In this paper we present the photometric and spectroscopic followup of ASASSN-13db during the 2014-2017 outburst. 
In Section \ref{datared} we describe the observations and the data reduction. 
In Sections \ref{photo-ana} and \ref{spectra-ana} we analyze the light curve and the spectral emission and absorption 
features observed during the outburst. In Section \ref{discussion} we discuss the nature of the outburst. 
Finally, in Section \ref{conclu} we summarize our results.


\section{Observations and data reduction \label{datared}}

\subsection{Photometry \label{photometry}}

ASASSN-13db was tracked during the outburst and return to quiescence by
the All Sky Automated Survey for SuperNovae \citep[ASAS-SN;][]{shappee14asas}, the Las Cumbres Observatory Global Telescope Network \citep[LCOGT;][]{brown13},
the Beacon Observatory telescope in Kent, and some amateur
astronomers.

\begin{figure*}
\centering
\includegraphics[width=15cm]{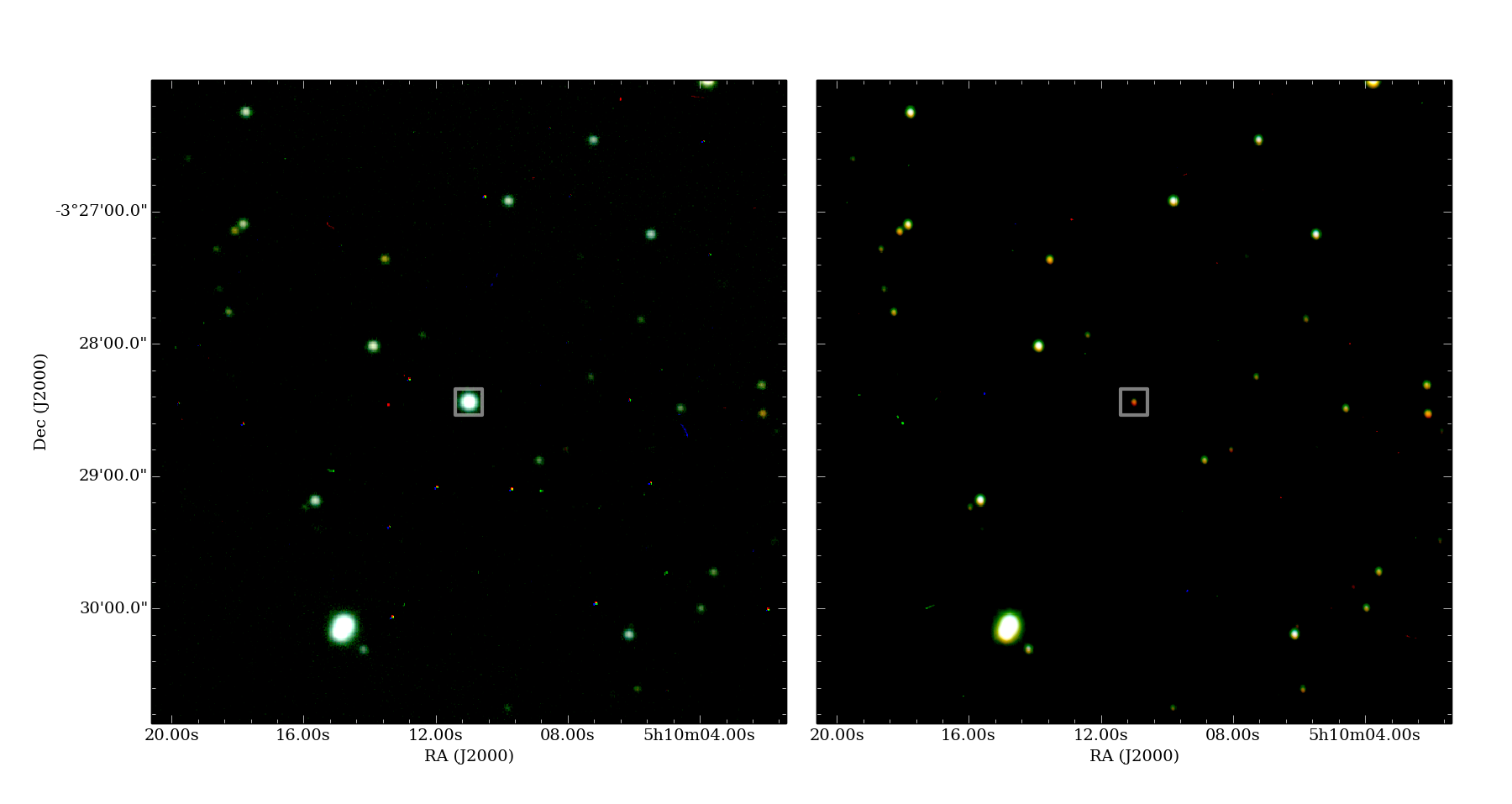}
\caption{LCOGT RGB (R=$i'$, G=$r'$, B=$g'$) images of ASASSN-13db obtained during the outburst (left, JD=2457110.243)
and post-outburst phases (right, JD=2457805.288). 
ASASSN-13db is marked with a square. We note the change in brightness and color.}
\label{LCOGTrgb-fig}%
\end{figure*}

The most complete photometric followup of the object since its discovery in 2013 is the V band light curve 
provided by ASAS-SN.  The ASAS-SN data were reduced using the standard ASAS-SN pipeline (Shappee et al. in prep.).  
We performed aperture photometry at the location of ASASSN-13db using the IRAF\footnote{IRAF is distributed by 
the National Optical Astronomy Observatory, which is operated by the Association of Universities for 
Research in Astronomy (AURA) under a cooperative agreement with the National Science Foundation. } \textit{apphot} 
package and calibrated 
the results using the AAVSO Photometric All Sky Survey \citep{henden14}.  Photometry and 5 $\sigma$ upper 
limits are reported in Table \ref{asasphoto-table}. The star is too dim for ASAS-SN
during quiescence. 
Two well-defined outbursts are seen in the data (see Figure \ref{ASAScurve-fig}). 
The first one, ending in January 2014, corresponds to a typical EXor outburst \citep{holoien14}. The
object increased in brightness by at least three magnitudes and returned to quiescence within a few months. The second outburst
started in November 2014 (approximately, on Julian Date [JD] 2456560) with a rapid increase in brightness of over four magnitudes
with respect to the minimum
in 2013/4, and ended in February 2017. Except for two $\sim$3-month gaps due to object visibility, 
we have continuous coverage. 
From now on, we refer to the three parts of the outburst, separated by observation gaps, 
as "A",  "B", and "C" (see Figure \ref{ASAScurve-fig}). The
2014-2017 outburst spans  $\sim$770 days, or about
2 years and 1.5 months.
It is very unlikely that the star returned to quiescence during the times when no observations are available, 
taking into account the timescale of the 2013 outburst and the length of the final dimming. We thus consider that the star has been in continuous outburst from November 2014
until February 2017. 

\begin{table}
\caption{ASAS-SN photometry. \label{asasphoto-table}}
\centering
\begin{tabular}{cc}
\hline \hline
JD & V  \\
(d) & (mag) \\
\hline
2456539.06603 & $>$16.93 \\
2456540.05305 & $>$16.86 \\
2456540.05443 & $>$16.94 \\
2456544.04266 & $>$16.80 \\
2456544.04403 & $>$16.83 \\
2456550.01831 & 16.37$\pm$0.17 \\
2456550.01967 & 16.64$\pm$0.18 \\
2456555.00243 & $>$15.64 \\
2456555.00386 & $>$15.71 \\
2456556.04073 & 15.62$\pm$0.14 \\
2456556.04209 & 15.79$\pm$0.22 \\
2456557.02096 & 15.56$\pm$0.15 \\
2456557.02233 & 15.44$\pm$0.16 \\
2456558.04729 & 15.49$\pm$0.12 \\
2456560.00528 & 15.38$\pm$0.13 \\
2456560.00666 & 15.25$\pm$0.12 \\
2456562.98509 & 14.53$\pm$0.04 \\
2456562.98645 & 14.45$\pm$0.04 \\
2456564.13425 & 13.91$\pm$0.02 \\
2456564.13562 & 13.88$\pm$0.02 \\
\hline
\end{tabular}
\tablefoot{Only the data at the beginning of the outburst are shown here. Upper limits correspond
to 5$\sigma$.
The complete photometry data is available online via the Centre de Donn\'{e}es astronomiques de Strasbourg, CDS.}
\end{table}

The multi-band photometric data from LCOGT were taken during the outburst (March 24-29, 2015) and
post-outburst/quiescence phases (February 2, 2017  - March 17,  2017), 
using the $u'$, $g'$, $r'$ and $i'$ Sloan filters. During the outburst, we obtained 100s of exposure per band. 
During the post-outburst phase, we obtained 
a short (60s) and a long (300s) exposure for $g'$, and
long (300s) exposures for $r'$ and $i'$. 
The object was too faint to be detected in $u'$ band during the post-outburst phase. 
Figure \ref{LCOGTrgb-fig} captures the impressive change in brightness and color
between the outburst and quiescence phases, as seen by LCOGT.

\begin{figure}
\centering
\includegraphics[width=9.5cm]{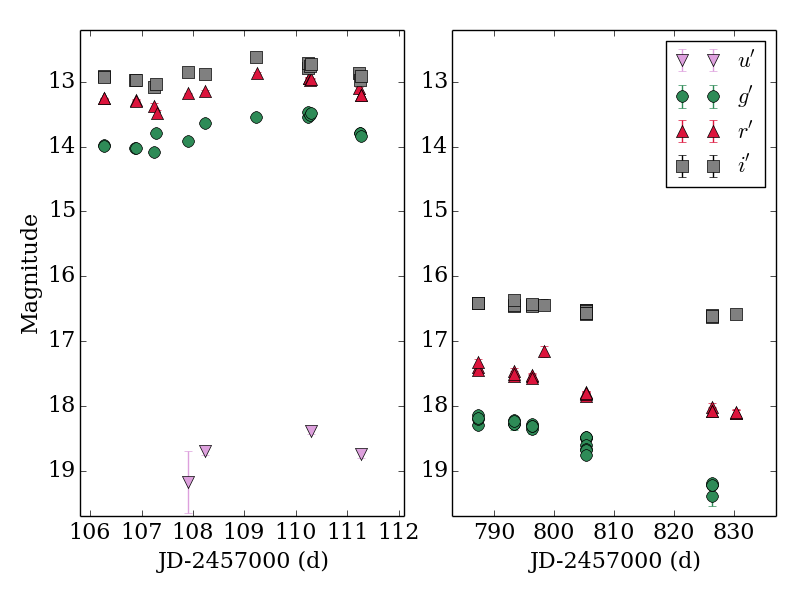}
\caption{LCOGT lightcurve in $u'$, $g'$, $r'$, and $'i$ filters. 
Given the long time span
between the two datasets, the outburst and post-outburst phases have
been separated. The magnitudes are calibrated using SDSS data. 
We highlight the change in color of the object in the post-outburst phase.}
\label{LCOGTcurve-fig}%
\end{figure}

The data were reduced (debiased, flat fielded, and aligned) using the standard LCOGT pipeline.
We then performed aperture photometry using IRAF \textit{daofind} and \textit{apphot} packages. Finally, we extracted the
relative instrumental magnitudes by calibrating all the observations against a LCOGT reference image 
(the best-quality data) and Sloan Digital Sky Survey \citep{gunn06,doi10};
data available via SkyServer\footnote{http://skyserver.sdss.org}. 
A total of 492, 738, and 640 stars were used to calibrate the $g'$, $r'$, and $i'$ filters, respectively.
We also attempted to calibrate the $u'$ band using the data from the outburst taken on JD 2457110.287, but 
due to having only five matches, of which at least two appeared to be variable, the calibration has a large error and
has to be regarded with extreme care.
The errors in the flux calibration of the LCOGT data (5\% for $g'$, 3\% for $r'$, and 4\% for $i'$, about 60\% for $u'$) 
have not been added to the total and thus do not appear in the table or plots. 
Table \ref{LCOGTphotometry-table} contains the final magnitudes from the LCOGT, and the resulting 
light curve is displayed in Figure \ref{LCOGTcurve-fig}.

\begin{table*}
\footnotesize{
\caption{\label{LCOGTphotometry-table} Final calibrated photometry from the LCOGT data. }
\centering
\begin{tabular}{lcccccccc}
\hline\hline
JD$_{ini}$ & JD$_{u'}$ &  $u'$ & JD$_{g'}$ & $g'$ & JD$_{r'}$ & $r'$ & JD$_{i'}$ & $i'$  \\
 &   & (mag) &   & (mag) &   & (mag) &   & (mag)  \\
\hline
{\bf Outburst}\\
2457106.26 & --- & --- & 2457106.258 & 13.97$\pm$0.01 & 2457106.263 & 13.26$\pm$0.01 & 2457106.261 &  12.91$\pm$0.01 \\
2457106.88 & --- & --- & 2457106.878 & 14.02$\pm$0.02 & 2457106.882 & 13.28$\pm$0.02 & 2457106.880 &  12.97$\pm$0.02 \\
2457106.89 & --- & --- & 2457106.888 & 14.02$\pm$0.02 & 2457106.892 & 13.30$\pm$0.02 & 2457106.890 &  12.97$\pm$0.02 \\
2457107.23 & --- & --- & 2457107.233 & 14.08$\pm$0.03 & 2457107.238 & 13.37$\pm$0.04 & 2457107.236 &  13.08$\pm$0.03 \\
2457107.29 & --- & --- & 2457107.285 & 13.78$\pm$0.04 & 2457107.289 & 13.47$\pm$0.04 & 2457107.287 &  13.03$\pm$0.04 \\
2457107.90 & 2457107.902 & 19.17$\pm$0.48$^u$ & 22457107.897 & 13.90$\pm$0.02 & 2457107.900 & 13.18$\pm$0.01 & 2457107.898 &  12.85$\pm$0.02 \\
2457108.23 & 2457108.238 & 18.70$\pm$0.06$^u$ & 2457108.232 & 13.63$\pm$0.02 & 2457108.236 & 13.15$\pm$0.01 & 2457108.234 &  12.88$\pm$0.02 \\
2457109.23 & --- & --- & 2457109.231 & 13.55$\pm$0.02 & 2457109.235 & 12.86$\pm$0.01 & 2457109.233 &  12.62$\pm$0.02 \\
2457110.23 & --- & --- & 2457110.231 & 13.54$\pm$0.03 & --- & --- & 2457110.23 & 12.78$\pm$ 0.04 \\  
2457110.24 & --- & --- & 2457110.243 & 13.47$\pm$0.02 & 2457110.247 & 12.94$\pm$0.01 & 2457110.245 &  12.71$\pm$0.02 \\
2457110.27 & --- & --- & 2457110.273 & 13.51$\pm$0.02 & 2457110.277 & 12.97$\pm$0.04 & 2457110.275 &  12.76$\pm$0.03 \\
2457110.29 & 2457110.295 & 18.39$\pm$0.04$^u$ & 2457110.287& 13.48$\pm$0.01 & 2457110.292 & 12.95$\pm$0.01 & 2457110.289 & 12.72$\pm$0.01 \\
2457111.23 & --- & --- & --- & --- & 2457111.235 & 13.10$\pm$0.02 & 2457111.233 & 12.86$\pm$0.02  \\
2457111.25 & --- & --- & 2457111.253 & 13.79$\pm$0.02 & --- & --- & 2457111.259 & 12.97$\pm$0.02  \\
2457111.27 & 2457111.274 & 18.75$\pm$0.06$^u$ &2457111.268 & 13.83$\pm$0.02 & 2457111.272 & 13.20$\pm$0.02 & 2457111.270 & 12.91$\pm$0.02  \\
2457702.50$^a$ & --- & --- & --- & --- & 2457702.497 &  15.26$\pm$0.07 & 2457702.553 & 16.09$\pm$0.06 \\
{\bf Post-outburst}\\
2457787.30 & --- & --- & 2457787.304 & 18.30$\pm$0.03 & 2457787.328 & 17.39$\pm$0.05 & 2457787.316 &  16.41$\pm$0.02 \\
2457787.31 & --- & --- & 2457787.308 & 18.20$\pm$0.04 & 2457787.332 & 17.45$\pm$0.05 & 2457787.320 &  16.41$\pm$0.02 \\
2457787.31 & --- & --- & 2457787.312 & 18.19$\pm$0.05 & 2457787.336 & 17.32$\pm$0.05 & 2457787.324 &  16.41$\pm$0.02 \\
2457793.33 & --- &--- & 2457793.328 & 18.28$\pm$0.09 & --- & --- & --- & --- \\
2457793.33 & --- & --- & 2457793.330 & 18.22$\pm$0.04 & 2457793.354 & 17.53$\pm$0.04 & 2457793.342 &  16.46$\pm$0.02 \\
2457793.33 & --- & --- & 2457793.334 & 18.28$\pm$0.04 & 2457793.358 & 17.46$\pm$0.04 & 2457793.346 &  16.44$\pm$0.02 \\
2457793.34 & --- & --- & 2457793.338 & 18.23$\pm$0.04 & 2457793.362 & 17.51$\pm$0.04 & 2457793.350 &  16.36$\pm$0.02 \\
2457796.34 & --- &--- & 2457796.344 & 18.36$\pm$0.08 & --- & --- & --- & --- \\
2457796.35 & --- & --- & 2457796.345 & 18.27$\pm$0.04 & 2457796.369 & 17.54$\pm$0.04 & 2457796.357 &  16.45$\pm$0.02 \\
2457796.35 & --- & --- & 2457796.349 & 18.31$\pm$0.04 & 2457796.373 & 17.52$\pm$0.04 & 2457796.361 &  16.46$\pm$0.02 \\
2457796.35 & --- & --- & 2457796.353 & 18.32$\pm$0.04 & 2457796.377 & 17.57$\pm$0.04 & 2457796.365 &  16.44$\pm$0.02 \\
2457798.33$^a$ & --- &--- & --- & --- & 2457798.338 & 17.16$\pm$0.08 & 2457798.330 & 16.44$\pm$0.08 \\
2457805.29 & --- &--- & 2457805.287 & 18.47$\pm$0.04 & --- & --- & --- & --- \\
2457805.29 & --- & --- & 2457805.288 & 18.49$\pm$0.01 & 2457805.312 & 17.80$\pm$0.01 & 2457805.300 &  16.52$\pm$0.01 \\
2457805.29 & --- & --- & 2457805.292 & 18.49$\pm$0.02 & 2457805.316 & 17.79$\pm$0.02 & 2457805.304 &  16.53$\pm$0.01 \\
2457805.30 & --- & --- & 2457805.296 & 18.48$\pm$0.02 & 2457805.320 & 17.81$\pm$0.01 & 2457805.308 &  16.53$\pm$0.01 \\
2457805.34 & --- &--- & 2457805.338 & 18.60$\pm$0.06 & --- & --- & --- & --- \\
2457805.34 & --- & --- & 2457805.340 & 18.66$\pm$0.03 & 2457805.364 & 17.85$\pm$0.03 & 2457805.352 &  16.58$\pm$0.02 \\
2457805.34 & --- & --- & 2457805.344 & 18.69$\pm$0.03 & 2457805.368 & 17.82$\pm$0.03 & 2457805.356 &  16.56$\pm$0.02 \\
2457805.35 & --- & --- & 2457805.348 & 18.76$\pm$0.03 & 2457805.372 & 17.80$\pm$0.03 & 2457805.360 &  16.56$\pm$0.01 \\
2457826.28 & --- &--- & 2457826.276 & 19.39$\pm$0.15 & --- & --- & --- & --- \\
2457826.28 & --- & --- & 2457826.277 & 19.22$\pm$0.06 & 2457826.301 & 18.01$\pm$0.06 & 2457826.289 &  16.62$\pm$0.02 \\
2457826.28 & --- & --- & 2457826.281 & 19.19$\pm$0.06 & 2457826.305 & 18.07$\pm$0.06 & 2457826.293 &  16.60$\pm$0.02 \\
2457826.29 & --- & --- & 2457826.285 & 19.22$\pm$0.07 & 2457826.309 & 18.08$\pm$0.06 & 2457826.297 &  16.61$\pm$0.02 \\
2457830.25 & --- &--- & 2457830.248 & --- & 2457830.272 & 18.12$\pm$0.06 & 2457830.260 & --- \\
2457830.25 & --- &--- & 2457830.252 & --- & 2457830.276 & 18.10$\pm$0.03 & 2457830.264 & --- \\
2457830.26 & --- &--- & 2457830.256 & --- & 2457830.280 & 18.10$\pm$0.04 & 2457830.268 & 16.59$\pm$0.02 \\
\hline  
\end{tabular}
\tablefoot{The data have been calibrated against the
Sloan filters and do not include the flux calibration errors (5\% for $g'$, 3\% for $r'$, and 4\% for $i'$; see text). 
The $u'$ data has only an approximate calibration with
expected 60\% systematic errors and are thus labeled as $^u$ (see text). In addition
to the initial JD for each set of observations, we list the individual
JD to account for the small observing-time differences between the various filters. Amateur data taken with the Sloan filters
are marked as $^a$.}
}
\end{table*}

The object was also tracked at the Beacon Observatory, associated with the
University of Kent. The observatory is equipped with a 17" Astrograph and 4kx4k 
CCD with 0.956 arcsec/pixel. The filters are standard Johnson V, R$_c$, and I$_c$. 
The observations were taken on a fair-weather basis between November 2016 and January 2017.
Exposure times range from 2 to 48 minutes, depending on the brightness of the star and the weather
conditions. The data were corrected for bias and flat fielding, and an astrometric solution was obtained. 
Aperture photometry was calibrated relative to
the data from JD=2457717.59 (estimated to be the best
night in terms of weather and seeing),
following an iterative procedure \citep{sicilia08}.
Each filter was calibrated independently, and the final errors include photometry and relative
calibration errors. The relative calibrations are found to be very stable and do not show
any strong magnitude- or color dependency. No absolute
calibration was possible in this case due to the lack of reference data.

\begin{figure}
\centering
\includegraphics[width=9.0cm]{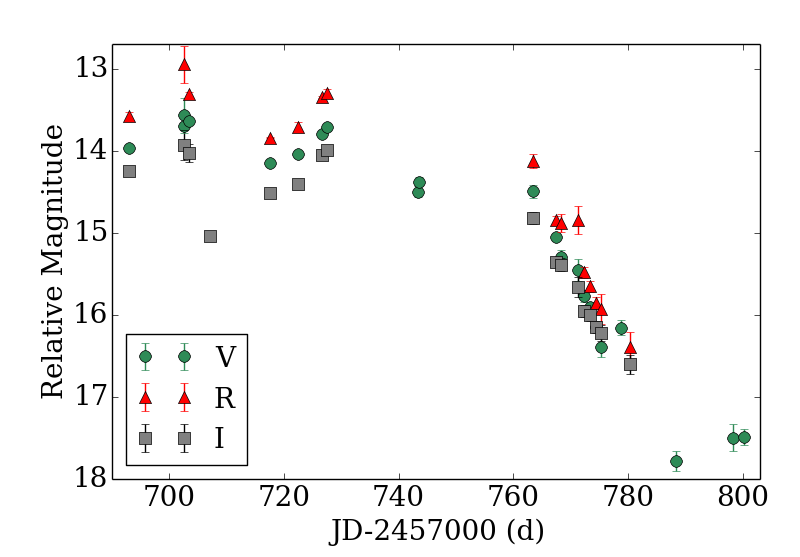}
\caption{Light curve from the Beacon Observatory. The magnitudes are relative to the data from
JD=2457717.594. The data show the rapid decrease
of the source flux during January and February 2017 and stabilization from February to March 2017.}
\label{kentlightcurve-fig}%
\end{figure}

Further data were provided by amateur astronomers R. Pickard and G. Piehler from the citizen science 
project HOYS-CAPS\footnote{Hunting Outbursting Young Stars with the Centre for 
Astropohysics and Planetary Science {\tt http://astro.kent.ac.uk/$\sim$df/hoyscaps/index.html }}
at the University of Kent.
These data come from various sources, including their own telescopes and further LCOGT data.
The data from Pickard were taken with the LCOGT 0.4, 1.0, and 2.0m telescopes. 
Thus, although most of the data were obtained for the VR$_c$I$_c$  and VRI Bessell filters,
there are some $r'$ and $i'$ data that are comparable to the rest of our LCOGT data, and thus
calibrated in the same way.
The data from Piehler were taken with a 510/2030mm Newtonic telescope with 
coma-corrector and a STL 11000M CCD camera. Data reduction was done with MAXIM 
DL 6.06. The filters used were a green TG filter and a clear filter CV. Although they are not
identical to the Johnsons V filter, they are similar enough and can be used to display the overall evolution
of the object during its return to quiescence. 
All the amateur data were calibrated against the best night observations from Beacon Observatory (or against
LCOGT data, for the $r'$ and $i'$ observations from Pickard),
and the results are fully consistent with them. Table \ref{kent-table} provides the results, which are
displayed in Figure \ref{kentlightcurve-fig}.

\begin{table}
\footnotesize{
\caption{\label{kent-table} Final calibrated photometry from the Beacon Observatory.}
\centering
\begin{tabular}{lccc}
\hline\hline
JD$_{ave}$ & V & R & I  \\
 &   (mag) &    (mag)  & (mag)   \\
\hline
2457693.042$^a$ & 13.97$\pm$0.03 & 13.58$\pm$0.05$^f$ & 14.24$\pm$0.05$^f$ \\
2457702.610 & 13.57$\pm$0.21$^b$ & 12.95$\pm$0.23$^b$ & 13.93$\pm$0.18$^b$ \\
2457702.548$^a$ & 13.70$\pm$0.03 & --- & --- \\
2457703.506 & 13.64$\pm$0.03 & 13.31$\pm$0.03 & 14.03$\pm$0.11 \\
2457707.143 & --- & --- & 15.04$\pm$0.06 \\
2457717.594$^*$ & 14.14$\pm$0.01 & 13.84$\pm$0.01 & 14.51$\pm$0.01 \\
2457722.452 & 14.04$\pm$0.05$^b$ & 13.71$\pm$0.05$^b$ & 14.41$\pm$0.05$^b$ \\
2457726.564 & 13.80$\pm$0.02 & 13.35$\pm$0.06 & 14.06$\pm$0.02 \\
2457727.543 & 13.71$\pm$0.04 & 13.29$\pm$0.05 & 14.00$\pm$0.06 \\
2457743.436$^a$ & 14.50$\pm$0.06$^f$ & ---  & --- \\
2457743.461$^a$ & 14.38$\pm$0.06$^f$ & ---  & --- \\
2457763.448 & 14.50$\pm$0.08 & 14.12$\pm$0.09 & 14.82$\pm$0.07 \\
2457767.441 & 15.05$\pm$0.05 & 14.85$\pm$0.06 & 15.35$\pm$0.05 \\
2457768.367 & 15.30$\pm$0.09 & 14.88$\pm$0.11 & 15.39$\pm$0.08 \\
2457771.329 & 15.45$\pm$0.13 & 14.85$\pm$0.17 & 15.66$\pm$0.12 \\
2457772.393 & 15.77$\pm$0.07 & 15.48$\pm$0.07 & 15.96$\pm$0.07 \\
2457773.375 & 15.90$\pm$0.06 & 15.65$\pm$0.06 & 16.01$\pm$0.07 \\
2457774.329 & 16.15$\pm$0.07 & 15.86$\pm$0.07 & 16.15$\pm$0.07 \\
2457775.330 & 16.39$\pm$0.13 & 15.93$\pm$0.18$^b$ & 16.22$\pm$0.10 \\
2457780.296 & --- & 16.39$\pm$0.19$^b$ & 16.60$\pm$0.12 \\
2457778.833$^a$ & 16.15$\pm$0.10$^f$ & --- & --- \\
2457788.294$^a$ & 17.78$\pm$0.13$^f$ & --- & --- \\
2457798.325$^a$ & 17.50$\pm$0.17 & --- & --- \\
2457800.275$^a$ & 17.49$\pm$0.10$^f$ & --- & --- \\
\hline  
\end{tabular}
}
\tablefoot{These
are instrument-relative magnitudes, compared to the magnitude observed on JD 2457717.59 (marked with $^*$). The average
JD is listed for each day of observations. Data taken by amateur observers are marked with $^a$. 
Data taken with filters other than the standard Johnsons/Cousins filters are marked with $^f$. 
Data affected by bad weather are marked with $^b$.}
\end{table}

\subsection{Spectroscopy \label{spectroscopy}}

\begin{table}
\caption{\label{spectralobs-table} Summary of spectroscopy observations. }
\centering
\begin{tabular}{lcccc}
\hline\hline
JD & Date & Instrument & V & Phase \\
   &      &            & (mag) &  \\
\hline
2456976.798 & 2014-11-15 & CHIRON & 13.7 & 0.168 \\
2456978.808 & 2014-11-17 & CHIRON & 14.0 & 0.653 \\
2456991.737 & 2014-11-30 & CHIRON & 14.0 & 0.771 \\
2456991.753 & 2014-11-30 & CHIRON & 14.0 & 0.775 \\
2456992.606 & 2014-12-01 & CHIRON & 13.5 & 0.981 \\
2456992.797 & 2014-12-01 & CHIRON & 13.5 & 0.027 \\  
2456978.794  & 2014-11-17 & FEROS & 14.0 & 0.649 \\
2456990.665$^{w}$  & 2014-11-29 & FEROS & 13.7 & 0.513 \\
2456995.700$^{w}$  & 2014-12-04 & FEROS & 13.8 & 0.727 \\
2457779.662 & 2017-01-26 & OSMOS & 16.5  & 0.816 \\
\hline  
\end{tabular}
\tablefoot{The V magnitudes listed are interpolations between
the two nearest observations (between 0.2 and 3 days apart), as there is no simultaneous photometry. The phase is calculated
assuming a period of 4.15d. The dates at which a  wind component is detected are marked by a $^{w}$.}
\end{table}

A total of nine spectra were taken during the outburst. Six of them were taken with CHIRON \citep{tokovinin13}, a highly stable cross-dispersed echelle
spectroscope deployed at the SMARTS 1.5m telescope\footnote{http://www.ctio.noao.edu/noao/content/chiron}. The remaining three
were taken using the Fiber-fed Extended-Range Optical Spectrograph \citep[FEROS;][]{kaufer99}, located at the European Southern Observatory/Max-Planck 
Gesellschaft (ESO/MPG) telescope in La Silla, Chile. Our CHIRON data have a resolution of R = 25,000  and a wavelength coverage from 
4200 to 8800 \AA. FEROS has a resolution of R = 48,000 and wavelength coverage $\sim$3700-9215 \AA\ 
\citep{kaufer00}. The coverage is not continuous, with FEROS having a gap at 
8860-8880 \AA, while the CHIRON data is distributed over 61 orders with gaps between most of them. The observations were 
performed during November-December 2014 (see Table \ref{spectralobs-table}).

The reduction of the 1800\,s-exposure FEROS spectra was performed using the FEROS 
pipeline, which involves de-biasing, flat fielding, extraction, and 
wavelength calibration. The CHIRON data were obtained in fiber mode and reduced with the CHIRON pipeline
\citep[e.g.,][]{buysschaert17}.
Due to the source being relatively faint for a 1.5m telescope, the CHIRON data are noisier than the FEROS spectra.
The emission lines observed in the high-resolution
data, identified using both CHIRON and FEROS datasets, are given in Section \ref{spectra-ana}.

The lines were classified using line lists observed in other young stars \citep{sicilia12,appenzeller86,hamann92}
and the National Insitute of Standards and Technology (NIST) 
database\footnote{http://physics.nist.gov/PhysRefData/ASD/lines\_form.html} for atomic spectra \citep{ralchenko10}.  
We excluded the parts of the spectrum affected by strong telluric emission and absorption features \citep{curcio64}, which
affects about 60 lines. A total of 31 lines were not found within the NIST database, and thus appear as `INDEF'. Among these, 
5 have also been observed in EX Lupi and likely correspond to strong transitions whose species
have not yet been identified.
The atomic constants of the lines (lower energy level E$_i$, upper energy level E$_k$ and transition probability A$_{ki}$)
were extracted from the NIST database. 
The complete line list is shown in 
Table \ref{alllines-table}. In total, we identify over 200 lines, about half of which are classified
as "strong". Although this number is lower than the number of lines cited by \citet{holoien14}, this is due to
the worse S/N of the high-resolution spectra. We estimate that a further $\sim$200 lines are present but hard to identify 
due to blends, S/N, and/or atmospheric contamination.

Towards the end of the outburst ( January 26, 2017), when the object had an approximate magnitude of V=16.5 mag, 
a 3$\times$600s further spectrum was obtained with the 2.4m Hiltner Telescope and the low-resolution spectrograph 
Ohio State Multi-Object Spectrograph \citep[OSMOS; R$\sim$1600;][]{martini11},
covering an approximate wavelength range between 3900 and 6800 \AA. The wavelength solution has shifts up to $\sim$2.7 \AA\, 
which results in some uncertainties in the line identification. Although the
object had nearly the same V magnitude as at the end of the 2013 outburst, the Hiltner spectrum is still dominated by
continuum and narrow emission lines, similar to the spectrum obtained during the 2013 outburst \citep{holoien14}
or during the small outbursts of EX Lupi \citep{herbig01,sicilia15}. The results from the low-resolution spectroscopy
are discussed in Section \ref{end-sect}.

\section{Basic properties during outburst and quiescence \label{photo-ana}}

\subsection{Basic properties of ASASSN-13db}

The basic properties of ASASSN-13db were revealed by \citet{holoien14}, 
using photometry and spectra taken during the quiescence phase after the 2013 outburst. The
star has a spectral type M5, which for its age and luminosity corresponds to a mass $\sim$0.15\,M$_\odot$ and
a radius $\sim$1.1\,R$_\odot$. We adopt these values throughout the paper, although our data suggest a slightly
lower radius (or luminosity) after the 2014-17 outburst (see Section \ref{mdot-sect}).

\begin{figure}
\begin{center}
\includegraphics[width=0.9\columnwidth]{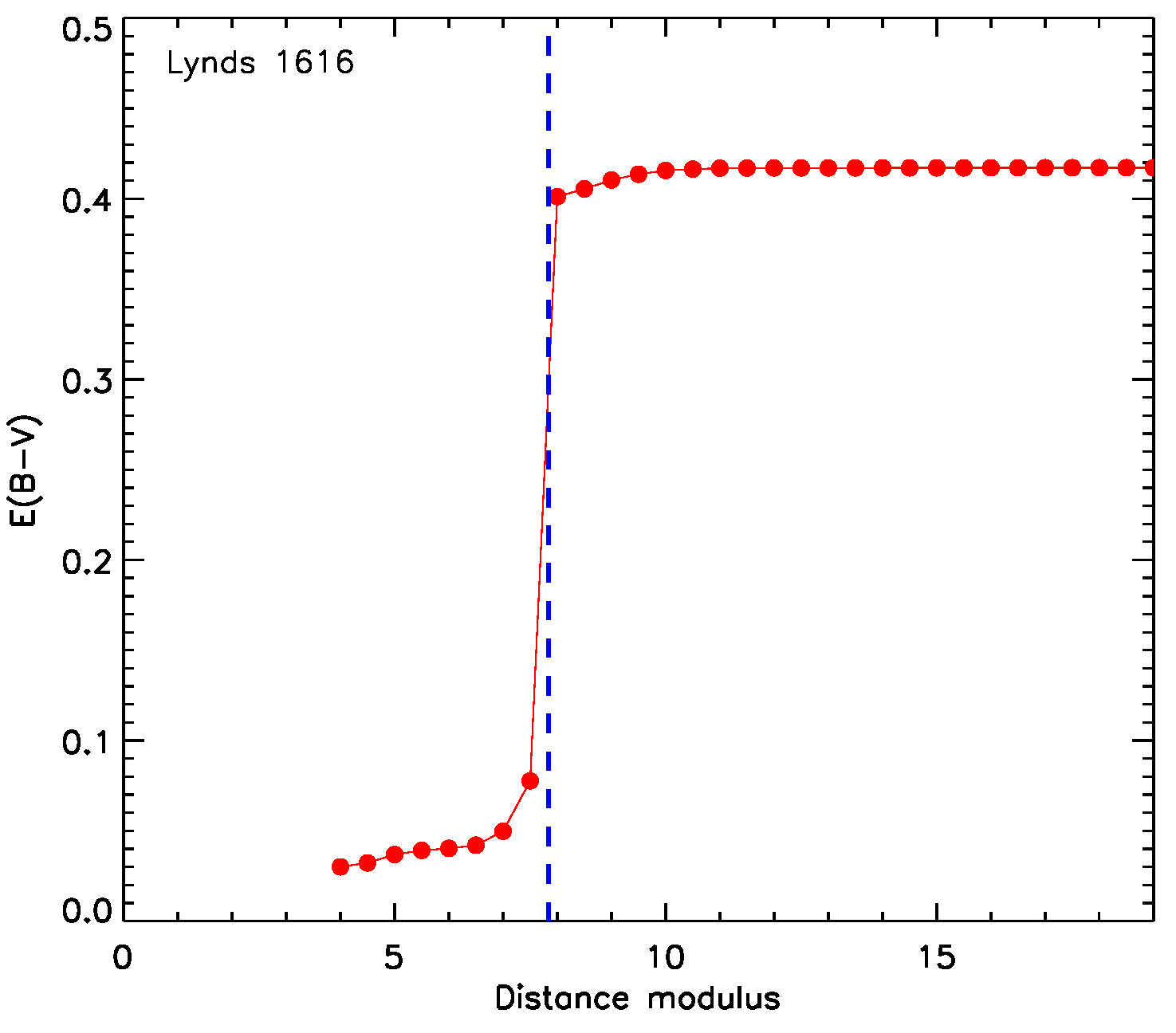}
\caption{The median cumulative reddening in each distance modulus bin  within  0.15$^{\circ}$ of the densest region in Lynds~1616,
using the data from \citet{green15}. The dashed line marks the most likely distance for Lynds~1616.}\label{Fig:ext}
\end{center}
\end{figure}

The distance to ASASSN-13db is likely similar to the distance to the Orion complex, 
usually estimated to be 400-450\,pc \citep[][]{jeffries07,reid09},
but more recently suggested to be $\sim$390 pc \citep{kounkel17}.
The distance of ASASSN-13db can be refined based on two arguments. First, the young star RX~J0510.3-0330,
at $\sim$2$'$ from ASASSN-13db, has a distance of 370$\pm$34\,pc estimated from GAIA \citep{gaia16,gaiabr16}. 
Second, the dark cloud Lynds~1616 is located near ASASSN-13db. We estimate the distance of Lynds~1616 
using the 3D extinction map from \citet{green15}. With 5-band $grizy$ Pan-STARRS~1 \citep{chambers16,flewelling16} 
photometry and 3-band 2MASS $JHK_{\rm s}$ photometry \citep{cutri03}
of stars embedded in the dust, \citet{green15} trace the extinction on $7'$ scales out to a distance of several kpc, 
by simutaneously inferring stellar distance, stellar type, and the reddening along the line of sight. 
Figure~\ref{Fig:ext} shows the median cumulative reddening in each distance modulus (DM) bin within  0.1$^{\circ}$ 
of the densest region of Lynds~1616. There is a rapid increase in the extinction at DM$\sim$7.5-8, 
suggesting a distance of 360$\pm$40\,pc. Assuming that both RX~J0510.3-0330 and 
Lynds~1616 belong to the same molecular cloud complex as ASASSN-13db, the distance of ASASSN-13db 
is likely $\sim$380\,pc.

\subsection{Characteristics of the outburst spectrum}

The spectra of classical T Tauri stars (CTTS) are characterized by numerous emission lines \citep{appenzeller86,hamann92}, 
with EXor variables being especially line-rich \citep{herbig01}.
The high number of metallic lines observed during the ASASSN-13db outbursts is remarkable 
and uncommon for CTTS, but similar to what has 
been observed in EX Lupi \citep{kospal08,sicilia12,holoien14}. 
The strongest line is H$\alpha$, with wings extending to $\pm$300 km/s, followed by other lines typical of 
accretion processes such as the Ca II IR triplet. Neutral and ionized Fe lines make up most of the emission spectrum. 
Most of the lines have low excitation potentials and correspond to features 
usually observed in absorption in the photospheres of late-type stars.
Their upper level energies, E$_k$, are in the range 2.4-6.7 eV, as observed in EX Lupi
\citep{sicilia12} and V1118 Ori \citep{giannini17} during outburst. 
Other neutral and ionized lines observed include Mg I/II, Ca I, Cr I, 
Co I, Ni I, Si II, V I, and Ti I/II.

Many lines have a strong, redshifted, absorption component with a width $\sim$100-200 km/s, together with a
blue-shifted emission component $\sim$50-150 km/s in width. Such profiles are classified as 
inverse P-Cygni or YY Ori-type profiles and are typical of systems viewed at high inclination angles (near edge-on), where the temperature
in the accretion flow decreases at larger distances from the star, and infalling matter is 
seen along the line-of-sight, leading to
higher-velocity material and potential obscurations of the star by the disk and the magnetosphere. 
The disk does not necessarily need to be edge-on with respect to the disk unless the accretion 
columns are polar, which may not be the case. For instance, in EX Lupi, the accretion spots in quiescence 
appear to hit the star at $\sim$50 degrees latitude \citep{sicilia15}, so for ASASSN-13db the angle may be high, but
the line-of-sight does not necessarily have to go through the disk or disk edge.

\onllongtab{
\begin{longtable}{lcccc}
\caption{ Emission lines observed during the outburst. \\ 
The lines are classified as ``strong" (S) or ``weak" (W). The ``References"
indicate whether the line has been observed in the 2013 outburst \citep[H14;][]{holoien14}, or in EX Lupi,
either in outburst \citep[SA12;][]{sicilia12} or quiescence \citep[SA15;][]{sicilia15}.
Uncertain classifications (due to several lines being consistent with the same wavelength, or to mismatches between
observed and potential laboratory wavelengths) are marked with `Unc.'. For some of these lines, a second possible
species may be responsible: these are given in the `Comments' column. Lines that appear blended with others
are labeled as `Ble.'. Lines that could not be associated with any
known transition are listed as `INDEF'. Lines affected by atmospheric contamination are marked with `Atm.', and INDEF
lines observed in EX Lupi are also labeled. Finally, lines for which the strong, wind-related blue-shifted absorption
is seen are marked with `BA'.  \label{alllines-table} }\\
\hline
\hline
Species & Wavelength (\AA) & Strength & References & Comments \\
\hline
\endfirsthead
\caption{Continued.}\\
\hline
Species & Wavelength (\AA) & Strength & References & Comments \\
\hline
\endhead
H I & 4861.28 & S &  SA12,H14,SA15  & H$\beta$ absorption \\
H I & 6562.57 & S &  SA12,H14,SA15  & BA \\
H I & 8413.32 & W &   SA12  &  \\
H I & 8437.95 & W &   SA12  &  \\
H I & 8467.26 & S &   SA12 & Ble. \\
H I & 8750.46 & W &  SA12   &  \\
H I & 9014.91 & S &  SA12  & Unc. Atm. \\
Ca I & 5041.62 & S &    &  \\
Ca I & 6450.86 & W &     & Unc. \\
Ca II & 8248.80 & W & SA15     &  \\
Ca II & 8498.02 & S & SA12,H14,SA15   &  \\
Ca II & 8542.09 & S & SA12,H14,SA15   &  \\
Ca II & 8662.14 & S &  SA12,H14,SA15  &  \\
Na I & 5889.95 & W &  SA12,H14,SA15   &  \\
Na I & 5895.63 & W &   SA12,H14,SA15  &  \\
C I & 9088.57 & S &   SA12 & Ble. Atm. \\
O I & 8446.25 & W &  SA12,SA15    & Unc. \\
Mg I & 5167.32 & S &  SA15  & BA \\
Mg I & 5172.68 & S &  SA15  &  \\
Mg I & 5183.6 & S & SA15   &  \\
Mg I & 5711.09 & W &     &  \\
Mg I & 8806.76 & S &    & Ble. \\
Mg II & 4390.59 & W &     &  \\
K I & 7664.90 & W &     & Atm. \\
S II & 6312.42 & W &     & Ble. \\
Si II & 5957.56 & S &  SA12  &  \\
Si II & 6347.10 & S &  SA12,SA15  &  \\
Si II & 6371.36 & W &  SA12,SA15   & \\
Ti I & 6312.03 & S &    &  \\
Ti I & 7069.07 & S &    & Unc. \\
Ti I & 8623.43 & S &    &  Atm. \\
Ti II & 4417.72 & W &   SA12,SA15  & Unc. \\
Ti II & 4468.50 & W & SA12,SA15    & Unc. Ble.  \\
Ti II & 4529.47 & S & SA12   &  \\
Ti II & 4571.98 & S & SA12,SA15   &  \\
Ti II & 5129.15 & S &  SA12  &  \\
Ti II & 5226.56 & S & SA12   &  \\
V I & 5727.78 & W &     & Unc.  \\
V I & 6324.66 & S & SA12   &  \\
V I & 8203.07 & S &    & Atm. Ble. \\
Co I & 7016.62 & S &  SA12  &  Unc.\\
Co I & 7085.10 & S &    & Ble.  \\
Cr I & 4077.09/.68 & W &  SA12   & Unc. \\
Cr I & 5204.52 & S &  SA12  &  \\
Cr I & 5208.44 & W & SA12    &  \\
Cr I & 5298.27 & S &  SA12  &  \\
Cr I & 6138.24 & S &    &  \\
Ni I & 5753.69 & S &    &  \\
Fe I & 4045.82 & W &  SA12,SA15   & Unc., BA \\
Fe I & 4063.55 & W &  SA12,SA15   & Unc., BA \\
Fe I & 4071.74 & W & SA12,SA15    & Unc. \\
Fe I & 4132.06 & W &  SA12,SA15   &  \\
Fe I & 4191.43 & W &  SA12,SA15   &  \\
Fe I & 4207.13 & W &   SA12  & Ble. \\
Fe I & 4208.60 & W &   SA12  & Ble. \\
Fe I & 4215.42 & W &  SA12   &  \\
Fe I & 4226.34 & W &  SA12   & Unc., BA \\
Fe I & 4258.61 & W &  SA12   & Unc. \\
Fe I & 4293.80 & S & SA12   & Ble. \\
Fe I & 4324.95 & W & SA12,SA15    &  \\
Fe I & 4325.74/.76 & W & SA12,SA15    &  \\
Fe I & 4375.93/.99 & S &  SA12,SA15  &  \\
Fe I & 4383.55 & S &  SA12,SA15  & BA \\
Fe I & 4389.24 & W &     &  \\
Fe I & 4408.41 & W &  SA12   &  \\
Fe I & 4415.12 & W &  SA12,SA15   &  \\
Fe I & 4461.65 & S &  SA12,SA15  &  \\
Fe I & 4482.17 & S & SA12,H14   &  \\
Fe I & 4494.46 & W &  SA12,H14,SA15   &  \\
Fe I & 4602.94 & S &  SA12  &  \\
Fe I & 4772.80 & S &  SA12  & Unc. \\
Fe I & 4939.24/.69 & S &  SA12  &  \\
Fe I & 4994.13 & S &  SA12  & Fe II 4993.36? \\
Fe I & 5012.07 & S &  SA12  &  \\
Fe I & 5041.07/.76 & S &  SA12  & Ca I 5041.62? \\
Fe I & 5051.63 & S &  SA12  &  \\
Fe I & 5060.03 & S & SA12   &  \\
Fe I & 5083.34 & S & SA12   &  \\
Fe I & 5110.36 & S &  SA12  &  \\
Fe I & 5123.72 & S &  SA12  &  \\
Fe I & 5151.90 & S &  SA12  &  \\
Fe I & 5227.15/.19 & S &  SA12  &  \\
Fe I & 5247.05 & S & SA12   &  \\
Fe I & 5250.65 & S &  SA12  &  \\
Fe I & 5254.95 & S & SA12   &  \\
Fe I & 5263.31 & S &  SA12  &  \\
Fe I & 5267.20 & W &     & Ble.  \\
Fe I & 5269.50/70.36 & S &  SA12,SA15  &  \\
Fe I & 5328.04/.53 & S &  SA12,SA15  &  \\
Fe I & 5332.90 & S &  SA12  &  \\
Fe I & 5341.02 & S &  SA12  &  \\
Fe I & 5371.49 & W & SA12,SA15    &  \\
Fe I & 5397.13 & S & SA12, SA15  &  \\
Fe I & 5455.61 & S &  SA12,SA15  &  \\
Fe I & 5497.52 & S &  SA12  &  \\
Fe I & 5501.47 & S &  SA12  & \\
Fe I & 5508.41 & S &  SA12  &  \\
Fe I & 5709.38 & W &   SA12  & Ble.  \\
Fe I & 5732.30 & W &  SA12   & Ble. \\
Fe I & 5753.12/.69/5.35 & S &  SA12  & Ble. \\
Fe I & 5816.06/.37 & S & SA12   & Ble. \\
Fe I & 5835.50/.57 & W &  SA12   & Ble.  \\
Fe I & 5853.68 & W &  SA12   &  \\
Fe I & 5859.61 & W &     &  \\
Fe I & 5862.36 & W &  SA12   &  \\
Fe I & 5916.25 & S &  SA12  & Ble.  \\
Fe I & 5958.33 & S &  SA12  &  \\
Fe I & 5976.78 & W &  SA12   & Ble.  \\
Fe I & 5984.81 & W &  SA12   & Ble. \\
Fe I & 6008.56 & W &  SA12   &  \\
Fe I & 6065.48 & S &  SA12  & Ble. \\
Fe I & 6116.04 & W &     &  \\
Fe I & 6127.91 & S &  SA12  & Ble. \\
Fe I & 6173.34 & W &  SA12   &  \\
Fe I & 6191.56 & S & SA12   & Ble. \\
Fe I & 6200.31 & S & SA12   &  \\
Fe I & 6230.72 & W &  SA12   &  \\
Fe I & 6232.64 & W &  SA12   & Ble. \\
Fe I & 6254.26/6.13 & S & SA12   & Blend  \\
Fe I & 6297.79 & S &  SA12  & Unc. Atm. \\
Fe I & 6318.02 & S & SA12   & Ble. \\
Fe I & 6336.82 & S &  SA12  &  \\
Fe I & 6393.60 & S & SA12   & Unc. \\
Fe I & 6400.00 & S &  SA12  & Unc. \\
Fe I & 6408.02 & S &  SA12  &  \\
Fe I & 6411.65 & S &  SA12  &  \\
Fe I & 6421.35 & S &  SA12  & Ble. \\
Fe I & 6498.94 & S & SA12   &  \\
Fe I & 6546.24 & S & SA12,H14   & Unc. \\
Fe I & 6574.10 & S & H14   & Ble.  \\
Fe I & 6609.11 & S & SA12,H14   &  \\
Fe I & 6705.12 & W &     & Ble.  \\
Fe I & 6750.15 & W &  SA12   & Unc. \\
Fe I & 6769.66 & S &    & Ble. \\
Fe I & 6841.34 & S &  SA12  & Ble. \\
Fe I & 6945.20 & S &  SA12  & Atm.  \\
Fe I & 6978.85 & S &  SA12  & Ble. \\
Fe I & 7024.06 & S & SA12   &  \\
Fe I & 7223.66 & S & SA12   & Ble. Atm. Unc. \\
Fe I & 7914.712 & S &    & Atm.\\
Fe I & 7937.14 & W & SA12    & Unc. \\
Fe I & 8027.94 & W &  SA12   & Atm.  \\
Fe I &  8048.99 & S &    & \\
Fe I & 8075.56 & S &    &  \\
Fe I & 8220.38 & W & SA12    & Ble., Atm.  \\
Fe I & 8327.06 & S &  SA12  & Atm., Unc. \\
Fe I & 8514.07 & S &    & Unc. \\
Fe I & 8582.26 & W &  SA12   &  \\
Fe I & 8824.22 & S &  SA12  &  \\
Fe I & 9147.71/.96 & S &    &  \\
Fe II & 4233.17 & S & SA12,SA15   & Unc.  \\
Fe II & 4273.33 & W & SA12,SA15    & BA  \\
Fe II & 4351.77 & W &  SA12,SA15   & Unc. \\
Fe II & 4489.18 & S & SA12,SA15   &  \\
Fe II & 4549.47 & S &  SA12,SA15  &  \\
Fe II & 4629.34 & S &  SA12,SA15  & Ti II 4629.47? \\
Fe II & 4666.76 & W &  SA12,SA15   &  \\
Fe II & 4731.45 & W &  SA12,SA15   & Var. \\
Fe II & 4923.92 & S &  SA12,SA15  & BA \\
Fe II & 5018.43 & S &  SA12,SA15  & BA \\
Fe II & 5080.31 & S &    & Fe I 5080.35? \\
Fe II & 5316.61 & S & SA12,SA15   &  \\
Fe II & 5534.85 & W &  SA12,SA15   &  \\
Fe II & 6113.32 & W &  SA12   &  \\
Fe II & 6247.56 & W &  SA12,SA15   &  \\
Fe II & 6432.68 & S & SA12   &  \\
Fe II & 6456.38 & W &  SA12,SA15   &  \\
Fe II & 6516.05 & S &  SA12,H14,SA15  &  \\
Fe II & 6678.88 & S &    &  \\
Fe II & 7111.71 & W &     & Unc.\\
Fe II & 7462.38 & S &  SA12  &  \\
Fe II & 7711.26/.71 & S & SA12   & Ble. \\
Fe II & 8839.06 & W &     &  Unc.  \\
INDEF & 4428 & S &    &  \\
INDEF & 5430 & S &    & \\
INDEF & 5498 & S &    & \\
INDEF & 5743 & W &     &  \\
INDEF & 5911 & S &    & FeI 5914?\\
INDEF & 5992 & S &  SA12  &  \\
INDEF & 6331 & W &     & FeI 6335? \\
INDEF & 6534 & W &     &  \\
INDEF & 6266 & S &  SA12  &  \\
INDEF & 6772 & S &    &  \\
INDEF & 6786 & W &     & Ble.  \\
INDEF & 6808 & W &     &  \\
INDEF & 6816 & W &     &  \\
INDEF & 6989/90 & S &    &  \\
INDEF & 7001 & S &    &  \\
INDEF & 7039 & S &    &  \\
INDEF & 7054 & W &     & Atm. \\
INDEF & 7197 & S &    & Atm.  \\
INDEF & 5395 & S &    & Ble., Fe I 5393? \\
INDEF & 5793 & W &  SA12   & \\
INDEF & 7724 & W &     &  \\
INDEF & 7749 & S & SA12   & Ble. \\
INDEF & 7790 & S &    &  \\
INDEF & 8366 & S &    &  \\
INDEF & 8389 & S &  SA12  &  Fe I 8387.71? \\
INDEF & 8403 & W &     &  \\
INDEF & 8676 & S &    &  \\
INDEF & 8758 & S &    & Atm.  \\
INDEF & 8945 & S &    & Ble. \\
INDEF & 8977 & W &     & Atm.  \\
INDEF & 9103 & W &     &  Atm.\\
INDEF & 9211 & S &    & He I? \\
\hline
\end{longtable}
}

\begin{figure*}
\centering
\begin{tabular}{ccc}
\includegraphics[width=5.7cm]{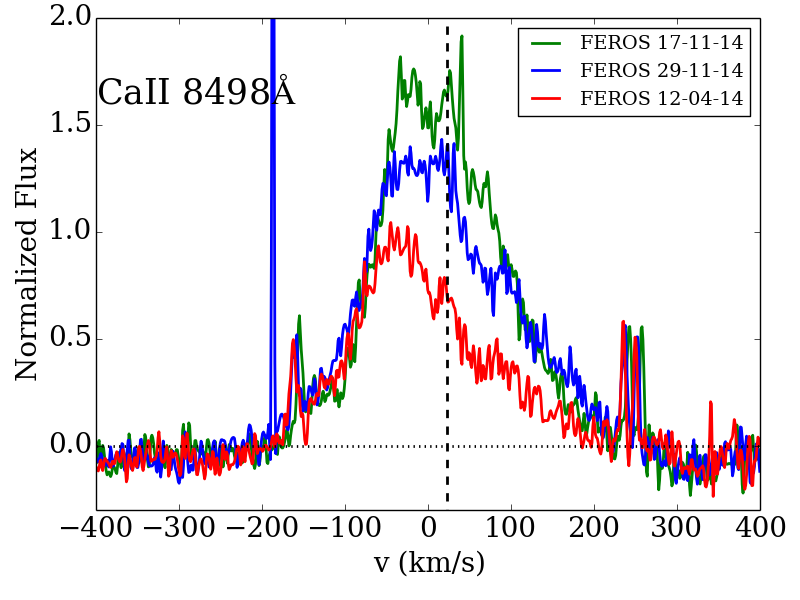} &
\includegraphics[width=5.7cm]{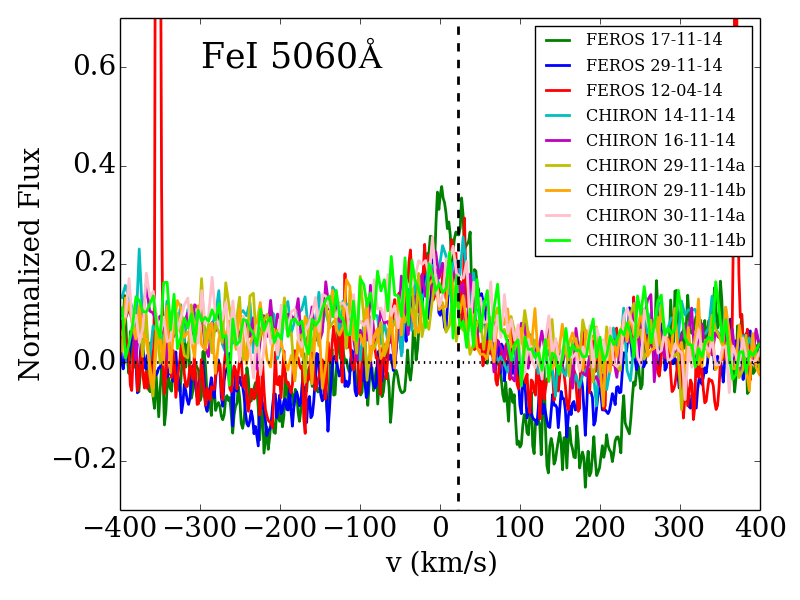} &
\includegraphics[width=5.7cm]{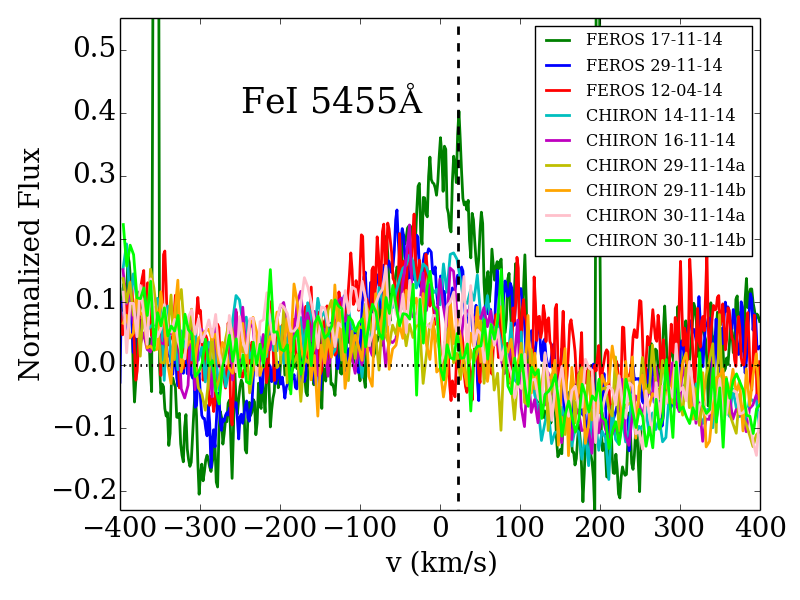} \\
\includegraphics[width=5.7cm]{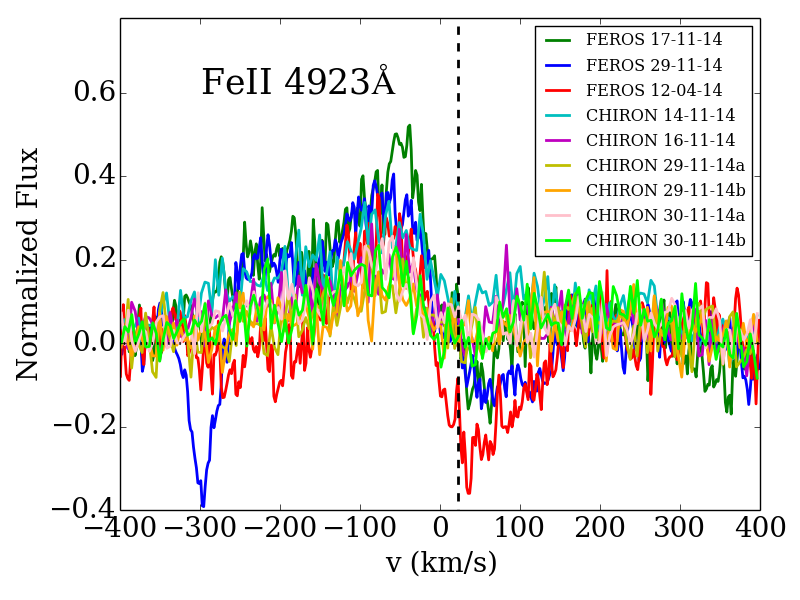} &
\includegraphics[width=5.7cm]{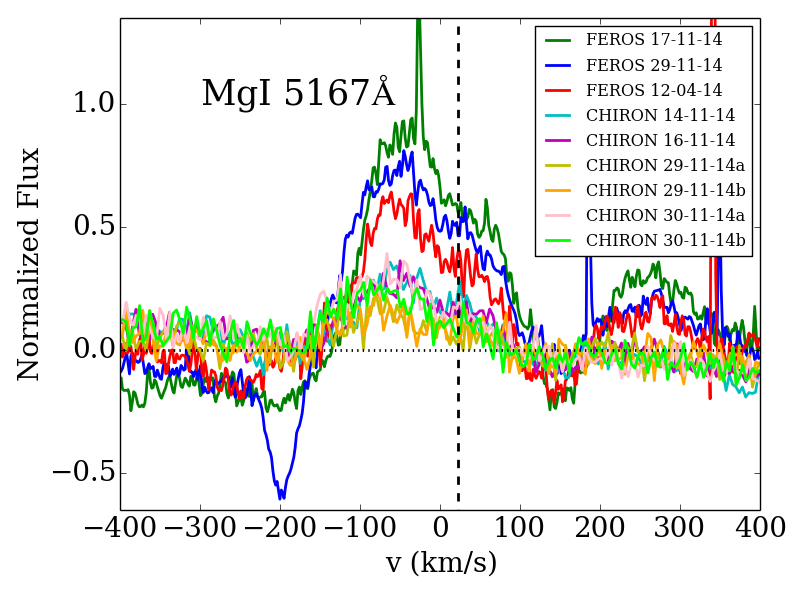} &
\includegraphics[width=5.7cm]{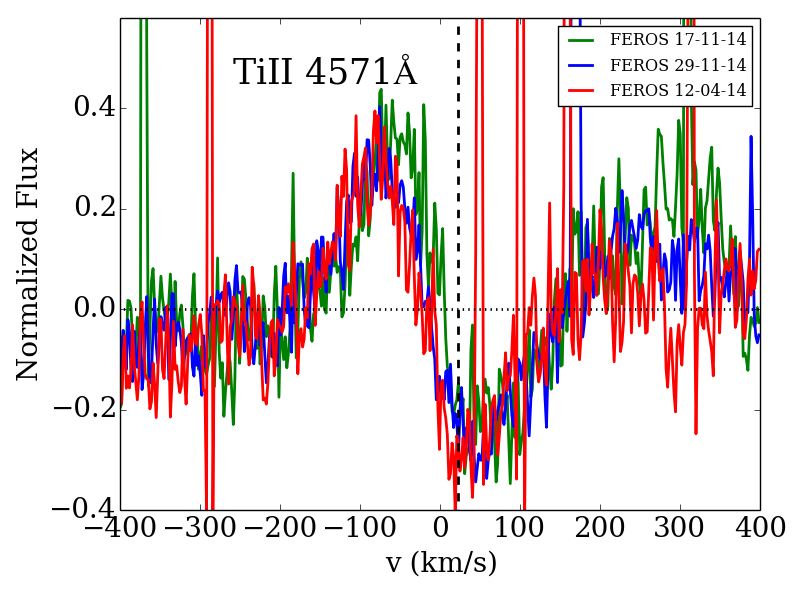} \\
\end{tabular}
\caption{Structure of several metallic neutral and ionized lines in the FEROS and CHIRON spectra during outburst.
The dashed line marks the radial velocity of ASASSN-13db.}
\label{lineexamples-fig}%
\end{figure*}

\begin{figure}
\centering
\begin{tabular}{c}
\includegraphics[width=6.5cm]{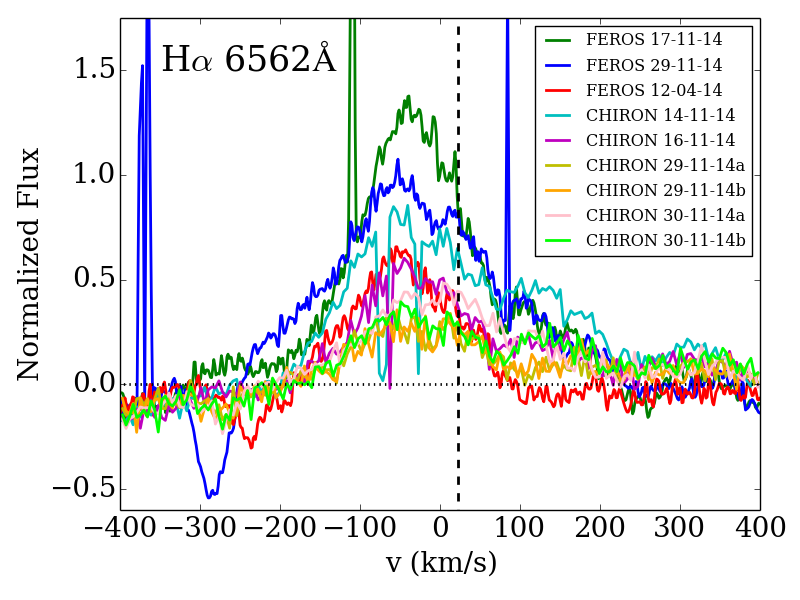} \\
\includegraphics[width=6.5cm]{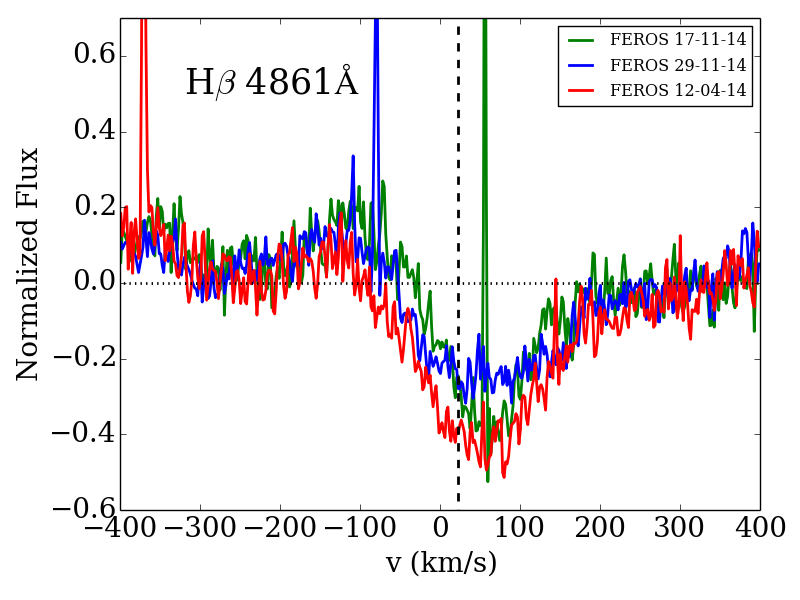} \\
\end{tabular}
\caption{Normalized and continuum-subtracted
H$\alpha$ (top) and H$\beta$ (bottom) lines as observed with FEROS and CHIRON (we note that H$\beta$ is only
covered by FEROS). While the H$\alpha$ line has a strong, asymmetric, emission component, H$\beta$ appears only in
absorption. The dashed line marks the radial velocity of ASASSN-13db. }
\label{halphahbeta-fig}%
\end{figure}

Ionized (Fe II, Ti II) lines correspond to relatively hot gas and are common in accreting, low-mass stars \citep[e.g.,][]{hamann92}.
Comparing to EX Lupi and other higher-mass EXors and accreting TTS, the most surprising characteristic of ASASSN-13db is the lack
of strong He I lines, which are usually among the strongest emission lines observed in CTTS \citep{hamann92}.
Like EX Lupi \citep{sicilia12}, ASASSN-13db shows no evidence for the forbidden lines common 
in CTTS \citep{hamann94}, which could indicate that there is no shock or that
the density of the surrounding material is high ($>$10$^5$ cm$^{-3}$) so that the shock is quenched \citep{nisini05}.
Examples of the typical velocity profiles are shown in Figure \ref{lineexamples-fig}, and Table \ref{stronglines-table}
lists the strongest lines observed.
Another interesting feature of ASASSN-13db is the absence of H$\beta$ emission. Although the H$\alpha$ line shows prominent
emission and a mild redshifted absorption asymmetry, H$\beta$ appears as a mildly redshifted
absorption feature (see Figure \ref{halphahbeta-fig}). The lack of He I and H$\beta$ emission could be a consequence of
the high inclination 
\citep[thus resulting in occultation of the hottest parts of the accretion shock, as observed in EX Lupi in outburst for high-energy lines;][]{sicilia15} or to 
low temperatures in the accretion structures associated with ASASSN-13db,
in contrast with solar-mass stars, as has also been suggested for brown dwarfs \citep{scholz09,bozhinova16a}. 
These possibilities are discussed in Section \ref{physicalproperties-sect}.

\begin{table}
\caption{\label{stronglines-table} Strong lines used to derive
the accretion properties.  }
\centering
\begin{tabular}{lccccl}
\hline\hline
Species & $\lambda$  & E$_i$ & E$_k$ & A$_{ki}$   & Notes \\
        & (\AA)      & (eV)  & (eV)  & (s$^{-1}$) & \\
\hline
Fe I & 4375.930 & 0 & 2.83 & 2.95E+04 & F \\
Fe I & 4461.652 & 0.08 & 2.87 & 2.95E+04 & F \\
Fe I  & 4482.169 & 0.11 & 2.88& 2.09E+04 &  \\
Ti II & 4571.980 & 1.57 & 4.28 & 1.92E+07 & F \\
Fe I & 4602.941 & 1.48 & 4.18 & 1.72E+05 &  F \\
Fe II & 4923.921 & 2.89 & 5.41 & 4.28E+06 & F, BA \\
Fe II & 5018.434 & 2.89 & 5.36 & 2.00E+06 & F, BA \\
Fe I & 5060.034 & 4.3 & 6.75 &  --- &  \\
Fe I & 5083.338 & 0.95 & 3.4 & 4.06E+04 &  \\
Fe I & 5110.358 & 3.57 & 6.0 & 9.99E+05 &  \\
Ti II & 5129.150 & 1.89 & 4.31 & 1.46E+06 & F \\
Mg I & 5167.32 & 1.48 & 3.88 & 2.72E+06 &  F, BA \\
Mg I & 5172.684 & 2.71 & 5.11 & 3.37E+07 &  F \\
Fe I & 5332.899 & 1.55 & 3.88 & 4.36E+04 &  \\
Fe I & 5341.024 & 1.6   & 3.93  & 5.21E+05 & \\
Fe I & 5397.127 & 0.91  & 3.21 &2.58E+04 &  \\
Fe I & 5455.609 & 1.01 & 3.28 & 6.05E+05 &  \\
Fe I & 5506.779 & 0.99 & 3.24 & 5.01E+04 & F \\
Fe I & 5916.247 & 2.45 & 4.55   & 2.15E+04 & \\
Si II & 5957.560 & 10.06 & 12.15 & 5.60E+07 &  \\
Fe I & 5958.333 & 2.17 & 4.26 &  --- &  \\
Fe II & 5991.376 & 3.15 & 5.22 & 4.20E+03 & F \\
Fe I & 6191.558 & 2.43 & 4.43 & 7.41E+05 & F \\
Fe I & 6200.312 & 2.61 & 4.61 & 9.06E+04 & F \\
Fe I & 6336.824 & 3.87 & 5.64 & 7.71E+06 &  \\
Si II & 6347.100 & 8.12 & 10.07 & 5.84E+07 &  \\
Fe I & 6393.601 & 2.43 & 4.37 & 4.81E+05 &  \\
Fe I & 6400.001 & 3.6 & 5.54 & 9.27E+06 & F \\
Fe I & 6421.351 & 2.28 & 4.21 & 3.04E+05 &  \\
Fe I & 6498.939 & 0.96  & 2.87 & 4.64E+02 & \\
H$\alpha$ & 6562.57 & 10.2 & 12.09 & 5.39E+07 & BA \\
Fe II & 6678.883 & 10.93 &      12.79 & 2.40E+07 & F  \\
Fe I & 6750.152 & 2.42  & 4.26 &1.17E+05 &  \\
Fe I & 6978.851 & 2.48 & 4.26 & 1.44E+05 &  \\
Fe I & 8048.990 & 4.14 & 5.68 &  --- &  \\
H I & 8467.26 & 12.1 & 13.55 &  3.44E+03 & \\
Ca II & 8498.020 & 1.69 & 3.15 & 1.11E+06 & F \\
Fe I & 8514.07  & 2.19  & 3.65 & 1.09E+05 & \\
Ca II & 8542.09 & 1.69  & 3.15 & 9.90E+06 & \\
Ca II & 8662.140 & 1.69 & 3.12 & 1.06E+07 &  \\
Mg I & 8806.756 & 4.36 & 5.75 & 1.27E+07 &  \\
Fe I & 8824.221 & 2.19 & 3.6 & 3.53E+05 & F \\
\hline  
\end{tabular}
\tablefoot{The `Notes' indicate whether a
blue-shifted absorption is present (BA) and whether the line was
fitted for the velocity analysis (F). }
\end{table}

\subsection{ Accretion rate estimates \label{mdot-sect}}

\begin{figure}
\centering
\includegraphics[width=8cm]{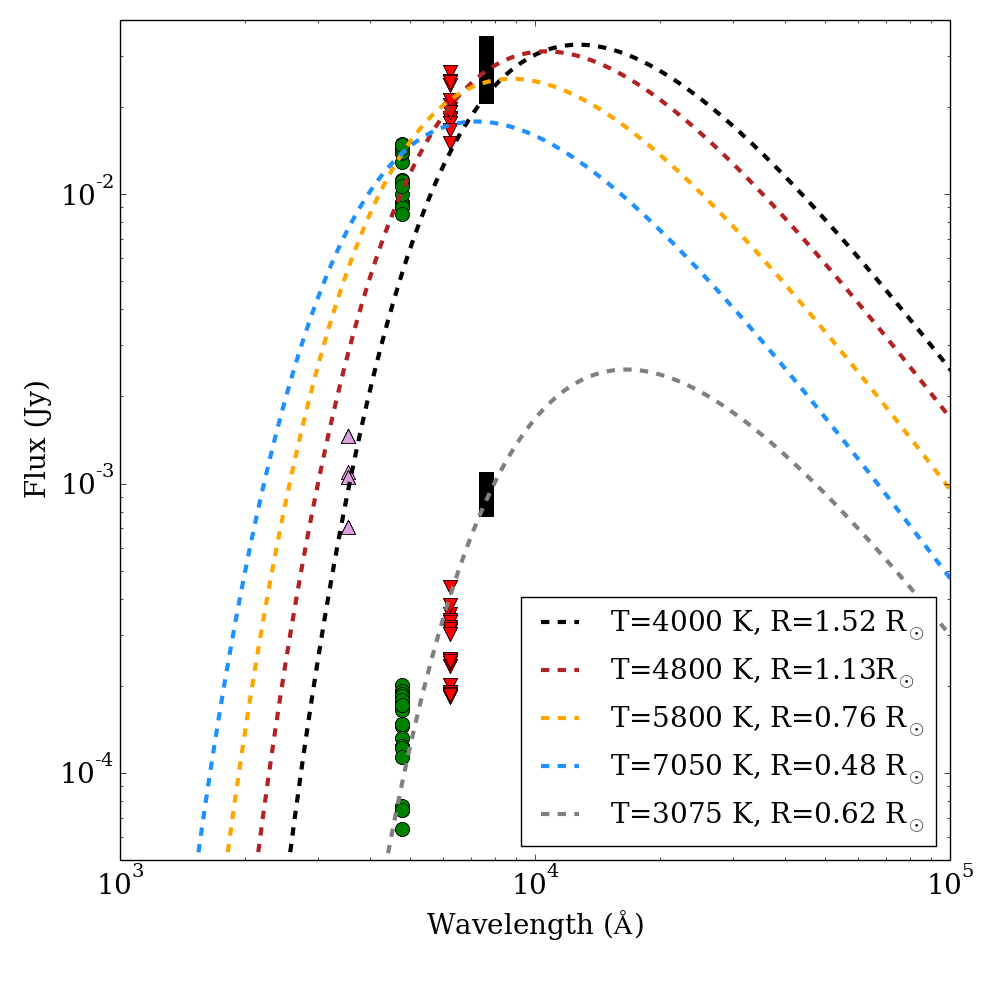} 
\caption{SED fits to the LCOGT data in outburst and quiescence. Black bodies with different temperatures and
effective radii are displayed for comparison.}
\label{sed-fig}%
\end{figure}

The best way to measure the accretion rate
during outburst is via the accretion luminosity. To estimate the total luminosity,
we assume that the bulk emission resembles that of a
star with an earlier spectral type. This is  usually the case for FUor objects \citep{hartmann96},
and it was observed for ASASSN-13db during the 2013 outburst \citep{holoien14}, for
EX Lupi \citep{juhasz12}, and many other EXors \citep{lorenzetti12}. We then integrate
the outburst spectral energy distribution (SED) observed by the LCOGT.
Given that the 2014-2017 outburst 
is stronger than the 2013 one, we take the 2013 outburst temperature \citep[4800K;][]{holoien14} as a lower limit.
Temperatures below $\sim$4800 K would result in either very unrealistic luminosities or extremely
large radii. Similarly, temperatures in excess of $\sim$7000 K result in effective radii that are too small, 
which gives us our upper limit. As seen in Figure \ref{sed-fig}, temperatures around 4800-5800 K produce the
best fits. Since we assume that extinction is negligible 
\citep[in agreement with][]{holoien14} and 
a black-body SED, the luminosities may be underestimated. Line emission may also affect the observed
magnitudes \citep{juhasz12}. The luminosities derived are around 0.5-0.6 L$_\odot$,
with the best fit being 0.6 L$_\odot$. After the 2017 outburst, the estimated luminosity drops to 0.03 L$_\odot$, which
is lower than the 0.06 L$_\odot$ measured by \citet{holoien14}. 
In general, the luminosity appears lower than the average for a M5 star \citep{fang17},
which may indicate a slightly later spectral type or variable extinction not accounted for at this stage
since the difference is minimal compared to the outburst luminosity. 
The luminosity obtained is very similar to the result derived from the ASAS-SN mean 
outburst magnitude (13.71$\pm$0.40 mag), assuming a distance of 380 pc and using the bolometric corrections for young
stars from \citet{kh95}.

The accretion luminosity can be translated into an accretion rate following
\citet{gullbring98},
\begin{equation}
L_{acc} = \frac{ G M_{star} \dot{M} }{R_{star}} \left( 1- \frac{R_{star}}{R_{infall}} \right), \label{lacc-eq}
\end{equation}
where G is the gravitational constant, \.{M} is the accretion rate, 
and R$_{infall}$ is the typical infall radius 
\citep[taken to be 5 R$_{star}$ by][]{gullbring98}\footnote{Note that, for a period 4.15d and M$_*$=0.15 M$_\odot$, 
the corotation radius is located
at approximately 9R$_*$ for a stellar radius R$_*$=0.62R$_\odot$, which would result in a slightly lower accretion rate.}.
From this, we can estimate an accretion rate in the range \.{M}=0.9-1.5 $\times$ 10$^{-7}$ M$_\odot$/yr,
or about \.{M}=0.7-3.3 $\times$ 10$^{-7}$ M$_\odot$/yr if we account for the observed span in magnitudes (Figure \ref{sed-fig})
and the uncertainty in the stellar
mass and radius. The same reasoning using the peak magnitude during
the 2013 outburst ($\sim$0.5 mag fainter) and an effective temperature of 4800 K
suggests that the accretion rate in 2013 was about half of the value observed in the 2014-2017 outburst.

\begin{figure*}
\centering
\begin{tabular}{ccc}
\includegraphics[height=3.7cm]{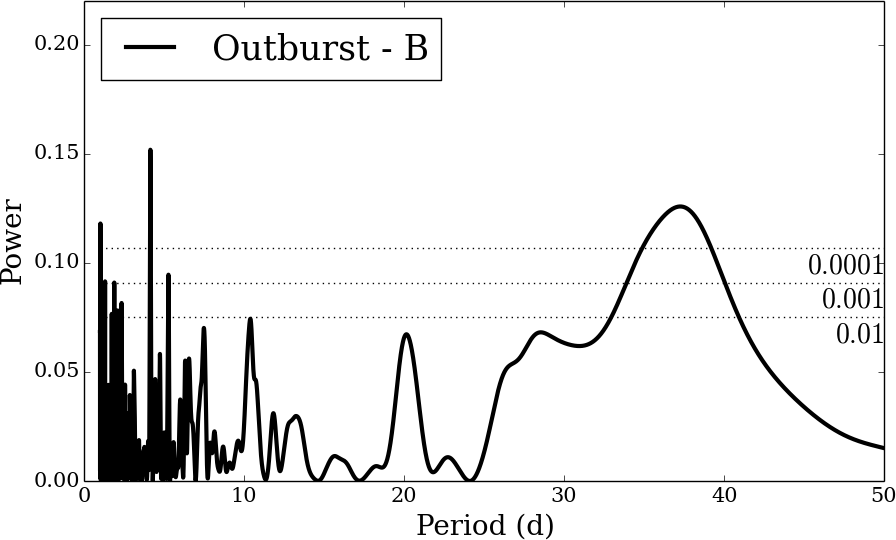} &
\includegraphics[height=3.7cm]{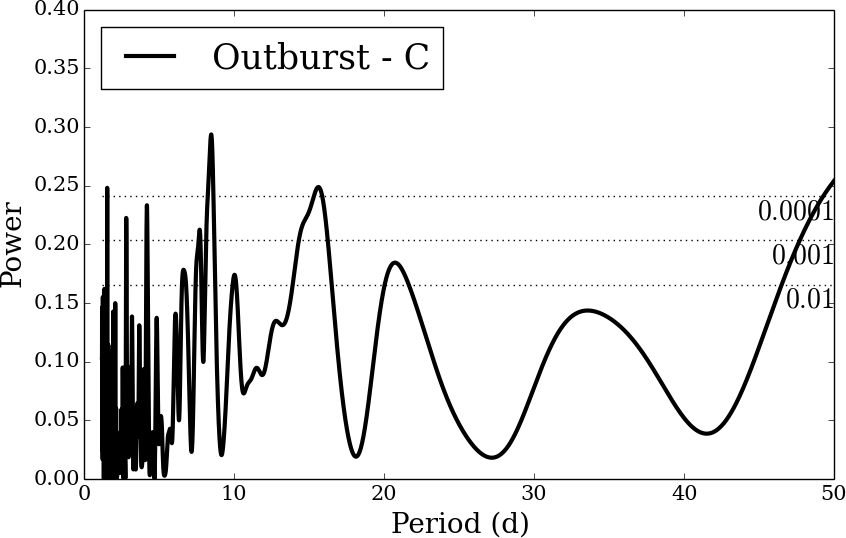} &
\includegraphics[height=3.7cm]{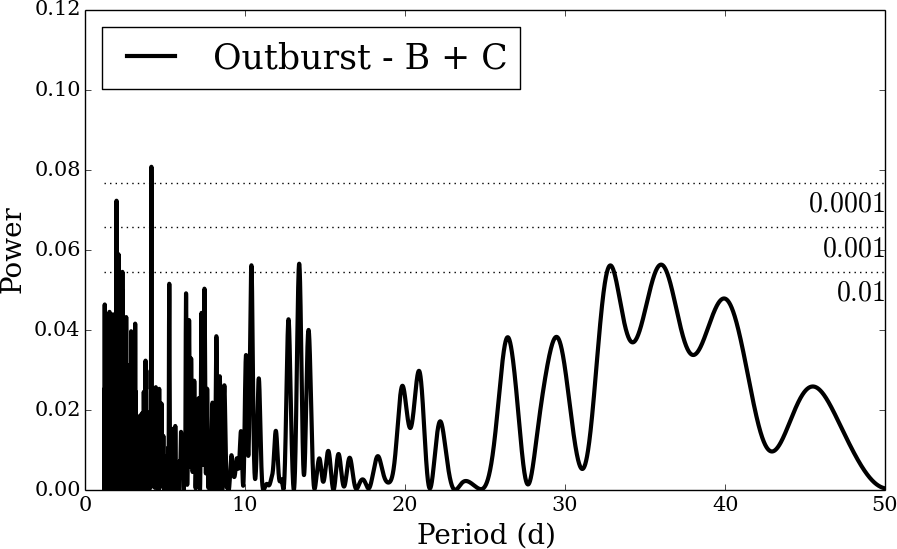}\\
\end{tabular}
\caption{GLSP periodograms for the 2014-2017 outburst during different epochs within
the outburst phase.}
\label{periodograms-fig}%
\end{figure*}

During the post-outburst phase, we
use the integrated line fluxes to derive the accretion rate from the flux-calibrated OSMOS spectrum.
Following the transformations between line luminosity and accretion luminosity from \citet{fang09}, 
we use the H$\alpha$, H$\beta$, and the He I 5875\AA\ line fluxes to derive the accretion luminosity,
and then Equation \ref{lacc-eq} to derive the accretion rate. The results
 are mutually consistent (Table \ref{Mdotquiescence-table}), with an average of
\.{M}$_{post-outburst}$=(1.2$\pm$0.2)$\times$10$^{-9}$ M$_{\odot}$/yr.
The uncertainty in this value could be of a factor of few, taking into account the assumptions for the
infall radius and the spread in the accretion rate versus line flux relations \citep{fang09}. This
accretion rate is on the high end for stars with similar stellar masses \citep{fang09,manara17}, 
but it may be enhanced with respect to the true quiescence value
since the underlying stellar spectrum is still heavily veiled.
In summary, ASASSN-13dbs experienced an increase in the accretion rate of over two
orders of magnitude between quiescence and outburst, which is similar to or higher than the increase in accretion
observed for EX Lupi in its 2008 outburst \citep{sicilia12,juhasz12}.

\begin{table}
\caption{\label{Mdotquiescence-table} Post-outburst line luminosities and accretion rates.}
\centering
\begin{tabular}{lccc}
\hline\hline
Line    & L$_{line}$    & L$_{acc}$     & \.{M}\\
        & (L$_\odot$) &   (L$_\odot$) &  (M$_\odot$/yr)  \\
\hline
H$\alpha$ & 2.0e-4 & 4.5e-3 & 1.4$\times$10$^{-9}$ \\
H$\beta$ & 5.3e-5 &  3.4e-3 & 1.0$\times$10$^{-9}$ \\
He I 5875\AA & 4.7e-6 & 4.3e-3 & 1.3$\times$10$^{-9}$ \\
\hline  
\end{tabular}
\end{table}

\subsection{Periodicities during the outburst phase}

\begin{figure}
\centering
\includegraphics[width=8.0cm]{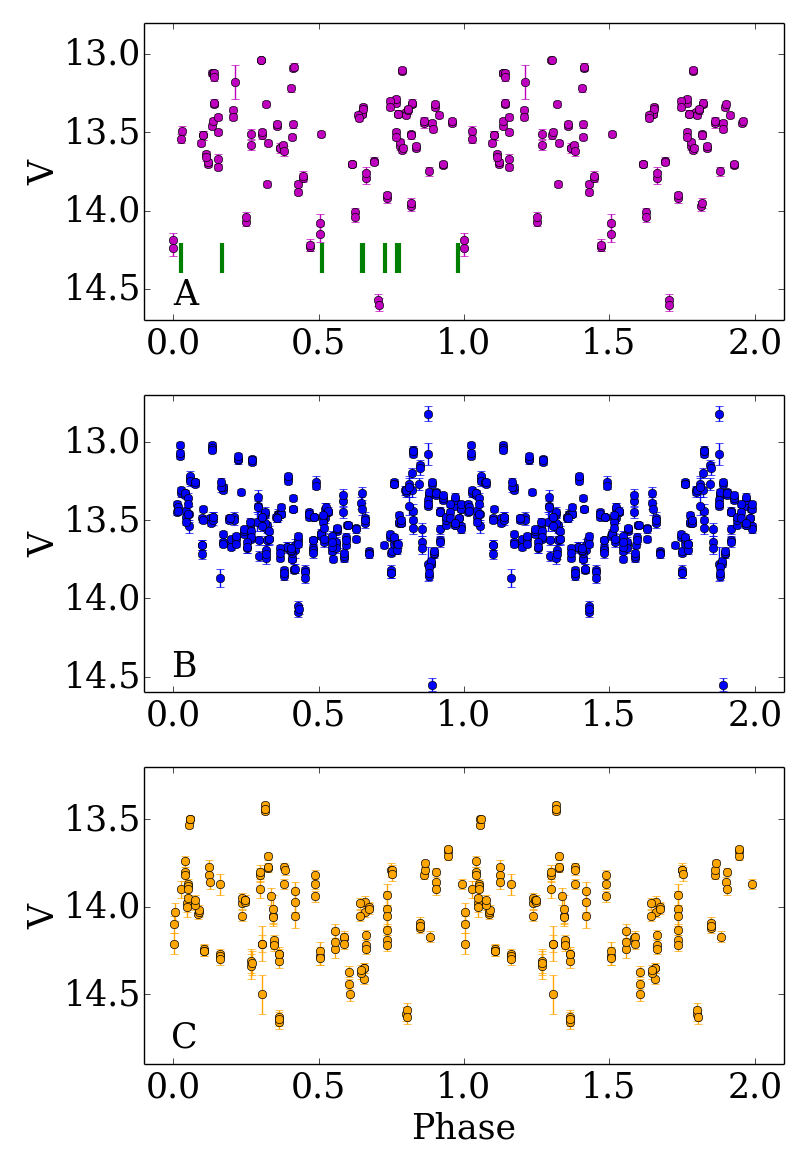}
\caption{Phase-folded ligth curve for the A, B, and C periods of ASAS-SN observations (see text)
assuming a 4.15$\pm$0.05\,d period. The data are repeated to display two phases. The phases at which
the FEROS and CHIRON spectra were obtained are marked as vertical green lines.}
\label{phasefolded-fig}%
\end{figure}

The temporal coverage of the ASAS-SN data allows us to search for periodicity in the 
light curve. For this, we run a generalized Lomb-Scargle periodogram \citep[GLSP;][]{scargle82,horne86,zechmeister09}.
Since the light curve is substantially different during the 2013 outburst, we restricted our study
to the 2014-2017 data. We then
explored the GLSP for the three parts of the outburst separated by observability gaps using the Python \textit{lomb\_scargle} routine.
Figure \ref{periodograms-fig} shows the periodograms taken during epochs B and C, when significant periodic signatures are detected.
The first dataset (A) shows
a quasi-periodic signal of about 12.6$\pm$0.6\,d, with several other peaks.
Dataset B reveals a strong periodic
signal (false-alarm probability, FAP$<$10$^{-7}$) at 4.15$\pm$0.05\,d, which also appears as a clear modulation
in the phase-folded magnitude (see Figure \ref{phasefolded-fig}), with an amplitude of about 0.5 magnitudes and a large
scatter. The same period, within errors, is also recovered in dataset C (4.20$\pm$0.09 d) and from the combination
of datasets B and C (4.15$\pm$0.03 d). Although the main peaks of the power spectrum in 
dataset A, A+B, or in the whole data are different, they all have peaks around 4.1-4.2 d.
The combination of all epochs has a potential periodic 
signal of 10.39$\pm$0.07\,d.
Both the 12.6 and the 10.4 d could be resonances or aliases of the 4.15\,d signature (3:1,5:2).
A similar phenomenon has been observed by \citet{mortier17}: signals at resonant periods tend to appear
when the number of stellar spots (which would correspond to accretion columns in our case) varies or when they
are evolving. It could be thus interpreted as changes in the accretion structures during
the beginning of the outburst, in epoch A, followed by a more stable accretion phase until the end of the outburst.
Folding the data by the longer periods does not lead to any conclusive results or shows quasi-periodic modulations
only in a small subset of data, and some FAP may be underestimated due to red noise. Table \ref{periods-table} summarizes the results.

\begin{figure}
\centering
\begin{tabular}{c}
\includegraphics[width=7.4cm]{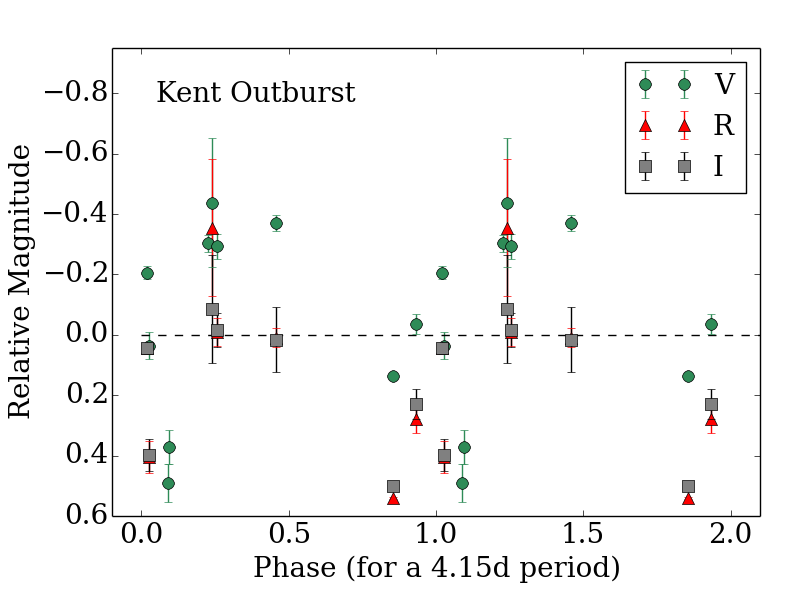}\\
\includegraphics[width=7.4cm]{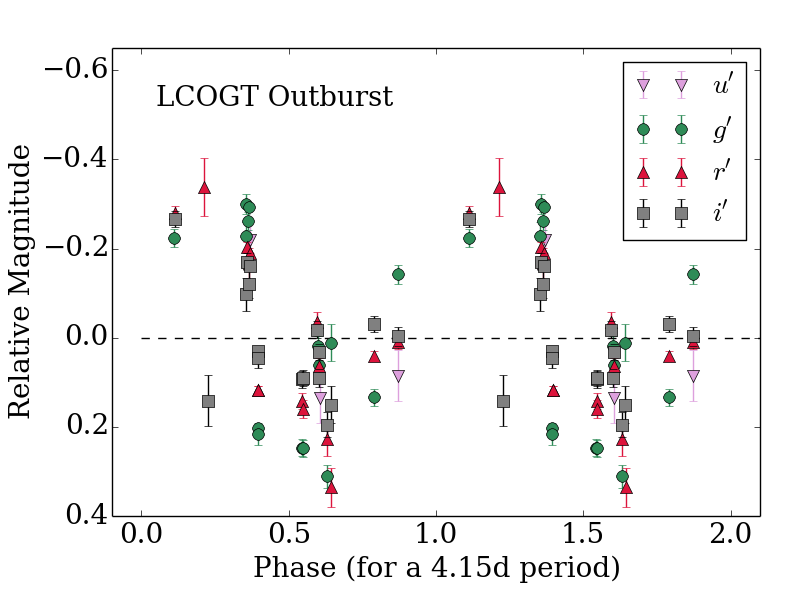}\\
\includegraphics[width=7.4cm]{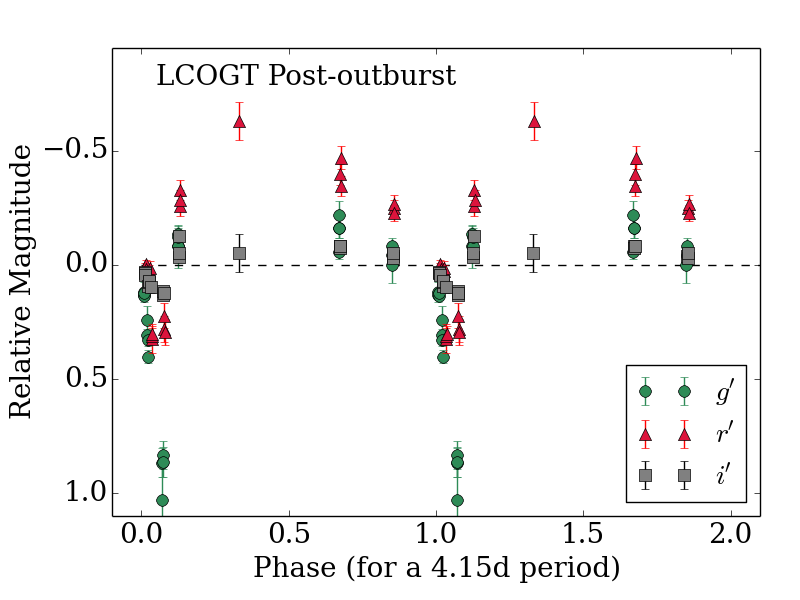}\\
\end{tabular}
\caption{Top: Phase-folded light curve for the data from the Beacon Observatory, using the
4.15d period.
Middle: Phase-folded light curve for the LCOGT data during outburst. 
Bottom: Phase-folded light curve for the LCOGT data during the post-outburst. 
All datasets are repeated to show two complete phases, and scaled to the average magnitudes. 
We note the phase shift between the datasets, which could be due to evolution of the spots.
The color change in the post-outburst phase may show a brief occultation event in the $g'$ band.  }
\label{phasefoldedLCOGT-fig}%
\end{figure}

The 4.15\,d period and the observed modulation are similar to those expected from rotational modulation of
a star with an asymmetric and slowly variable distribution of spots. Considering a M5 spectral type and
1-2 Myr age, we can expect a mass of approximately 0.15$\pm$0.05 M$_\odot$ according to 
\citet{siess00}. In this mass and age range, 4.15\,d is within typical rotation 
rates \citep{herbst02,lamm05,littlefair10,scholz13,bouvier14}. 
The modulations observed in the LCOGT and Kent data during outburst are also consistent with the
4.15d period (see Figure \ref{phasefoldedLCOGT-fig}),
although none of the datasets contain enough points to allow for an independent estimate of the period.
The fact that the shape and phase of the curve seems to change slightly with epoch may be due to 
slow changes in the spot distribution (possibly related to evolution of the accretion columns) throughout the outburst, although
the uncertainties in the period  over long time spans may also contribute.

\begin{table}
\caption{\label{periods-table} Observed periods during the 2014-2017 outburst and their false-alarm probabilities (FAP).
}
\centering
\begin{tabular}{lccl}
\hline\hline
Epochs & Period & FAP & Comments \\
      & (d)    &     &     \\
\hline
A       & 1.531$\pm$0.007 & $<$5e-5 & Alias?\\
"       & 12.66$\pm$0.61 &  $<$1e-4 & 3:1 Resonance? \\
"       & 2.856$\pm$0.029 &  $<$1e-4 & 2:3 Resonance? \\
B       & 4.146$\pm$0.045 & $<$1e-7 & Strong\\
C       & 8.42$\pm$0.43 &  $<$1e-4 & 2:1 Resonance? \\
"       & 4.200$\pm$0.095 &  $<$2e-3 & \\
A + B   & 4.141$\pm$0.015 &  $<$1e-6 & \\
"   & 10.376$\pm$0.084 &  $<$1e-6 & 5:2 Resonance? \\
B + C   & 4.147$\pm$0.028 &  $<$1e-4 & \\
All     & 12.63$\pm$0.10  & $<$1e-6 & 3:1 Resonance?  \\
"      & 10.386$\pm$0.066  & $<$1e-6 & 5:2 Quasi-periodic  \\
"        & 4.144$\pm$0.012   & $<$5e-6 &  \\
"        & 6.321$\pm$0.024  & $<$5e-6 & 3:2 Resonance? \\
\hline  
\end{tabular}
\tablefoot{For each date, the periods are arranged in order of likelihood.
The dates are defined as in Figure \ref{ASAScurve-fig}, with "All" including all three epochs A, B, and C.}
\end{table}

\subsection{The end of the outburst \label{end-sect}}

When the object became observable in September 2016, it had decreased in brightness by about 0.5 mag, compared to
epoch B of the outburst, and was slowly declining.
In December 2016, ASAS-SN and Beacon Observatory
data showed that the magnitude was rapidly decreasing. By
February 2017 the object had reached the previous quiescence levels, terminating the outburst after nearly 800 days.
The last results from the LCOGT (March 2017)
indicate that the object is still experiencing a very slow magnitude decrease and changes in 
color (see Figure \ref{LCOGTcurve-fig}), meaning that the steady state may not have been reached yet.


\begin{figure*}
\centering
\begin{tabular}{cc}
\includegraphics[width=8.cm]{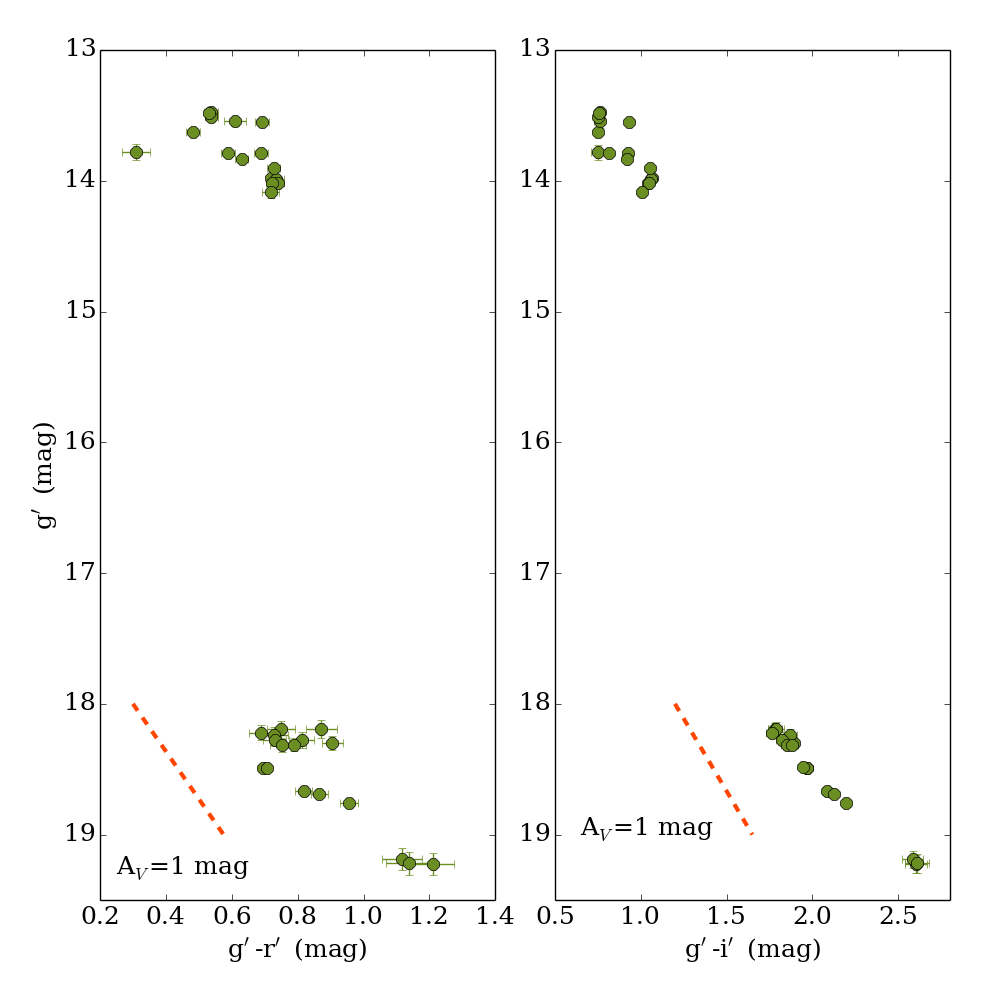} &
\includegraphics[width=8.cm]{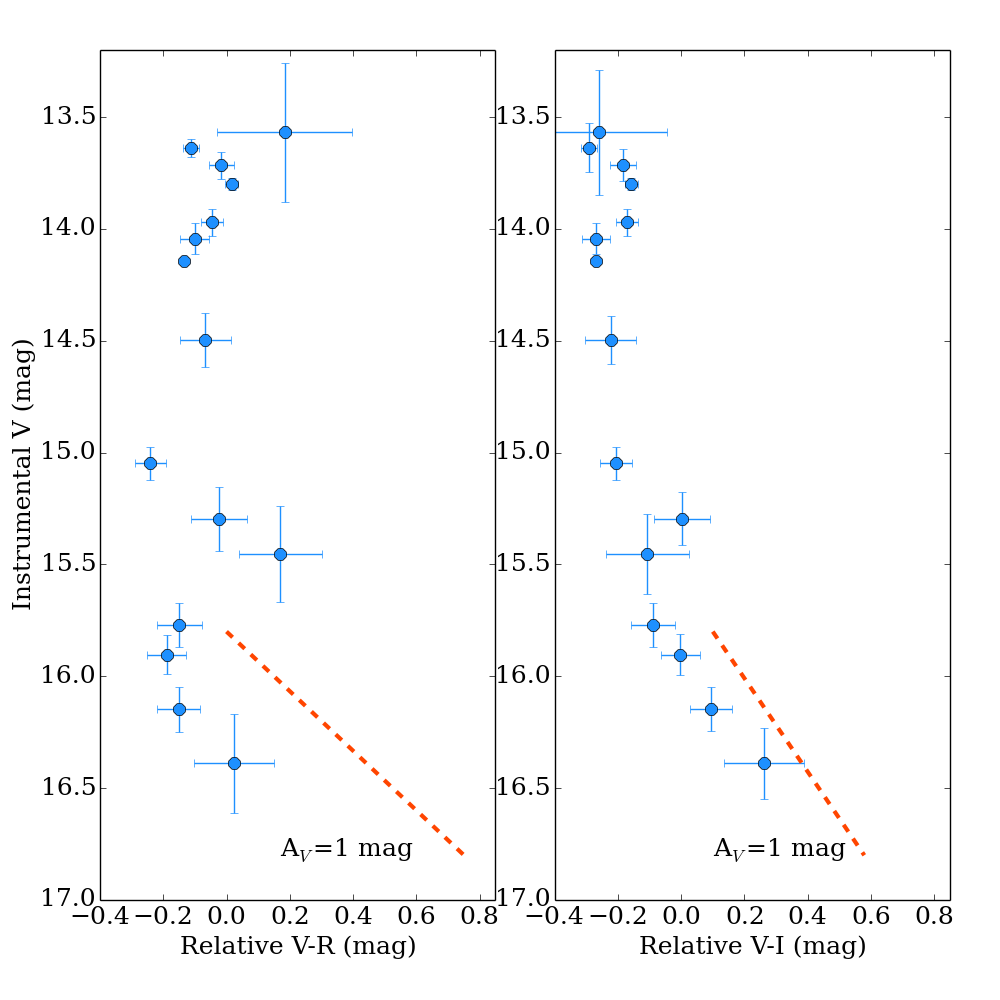}  \\
\end{tabular}
\caption{Color-magnitude diagrams showing the color evolution as observed by the LCOGT (left) and the Beacon Observatory (right). 
For the Kent data, since the different filters are not cross-calibrated, the colors are shifted around the median value and 
only the relative color variation is relevant. A reddening vector \citep{cardelli89,stoughton02} is shown for comparison.}
\label{colorevol-fig}%
\end{figure*}

The color evolution of ASASSN-13db can be studied with the LCOGT and Kent data (see Figure \ref{colorevol-fig}). The LCOGT
data covers part of the high and low states in all $g', r', i'$ bands, while the Kent data sample most of the
dimming with data taken between November  2016 and February 2017; in the last part of the outburst, however, we could only obtain
V data (thus no colors are available) at Kent. Although the Kent filters are not cross-calibrated,
the data provides valuable information about the dimming
process and the relative color evolution. The colors reveal that a luminosity change by approximately four magnitudes occurs at nearly constant color,
and thus cannot be attributed to extinction. The colorless magnitude change could be caused by an extended 
hot-spot region shrinking as the accretion rate decreases, followed by cooling down towards the M5 spectrum. 
Similar color changes, including a substantial colorless magnitude decrease, have been observed in other outbursting stars such
as V1118 Ori \citep{audard10}. At the last stage of the outburst, the object 
experiences a rapid color change, becoming redder faster than expected from extinction alone.
If the outburst phase is dominated by hot continuum ($>$4800 K) and the quiescence state is dominated by
the emission of a M5 star (T$_{eff}$=3075 K), the source should become significantly redder even in the absence of
extinction, with a color change of about 1 mag in V-R$_c$ and
about 2.5 mag in V-I$_c$ \citep[using the colors for young Taurus stars from][]{kh95}, which are larger than
observed.
The typical Sloan colors for M5 stars also suggest that the object will likely evolve to
redder magnitudes \citep[$g'-r'$=1.43 mag, $g'-i'$=3.21 mag for a M5 star according to][]{fang17}.
As of March 2017, it appears that the star is still dimming and evolving in color, and we predict that
it will likely be redder in quiescence.

\begin{figure*}
\centering
\includegraphics[width=19.5cm]{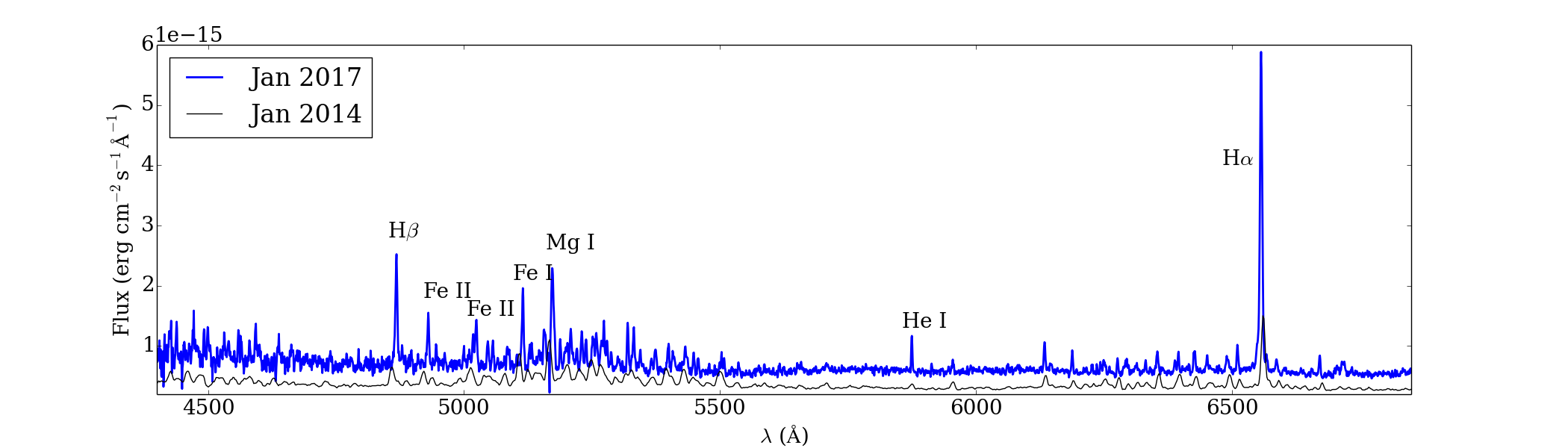}
\caption{ The post-outburst low-resolution OSMOS spectrum (blue), 
compared to the 2013 post-outburst data (black). 
Some of the strongest lines are labeled. We note that the stellar photospheric features were not visible as of January 2017.}
\label{osmos-fig}%
\end{figure*}

The OSMOS spectrum taken near the end of the outburst (Figure \ref{osmos-fig}) shows 
no evidence of self-absorption or complex line structure except in 
the H$\alpha$ line. Despite the low resolution ($\sim$50 kms$^{-1}$/pix),
the deep and broad absorption features observed in outburst would be still visible if they were present.
Identification of weak
or blended lines is highly uncertain due to the low resolution, 
so we list only the strongest lines (see Table \ref{osmos-table}).
H$\alpha$ is the only line that 
appears asymmetric, with a redshifted absorption that does not go
below the continuum level. H$\beta$ and He I 5875\AA\ are now detected in emission, suggesting that higher 
temperature regions become visible as the accretion column becomes less optically thick.
This was also observed in EX Lupi, where the higher-energy lines become visible during quiescence \citep{sicilia15}.
Although the object had faded back to V$\sim$16.6 mag when the 
spectrum was taken, the photosphere of the M5 star was not visible yet, and the spectrum still resembled those from the
2013 outburst with strong, relatively narrow, emission lines. 

\begin{table}
\caption{\label{osmos-table} Strong lines observed in the post-outburst phase. }
\centering
\begin{tabular}{lcl}
\hline\hline
Species & $\lambda$ & References \\
  & (\AA) & \\
 \hline
CrI: & 4078 & SA15 \\
FeII & 4351 & SA15 \\
H$\beta$ & 4861 & SA12/SA15 \\
FeI/FeII & 4939/4924 & SA12/SA15 \\
FeII & 5018 & SA12/SA15 \\
FeI & 5027 & SA12 \\
FeI & 5050 & SA12 \\
FeI & 5056 & SA12 \\
FeI: & 5110 & SA12 \\
MgI & 5173 & SA15/SA12 \\
HeI & 5876 & SA15/SA12 \\
FeI/CrI & 6137/6138 & --- \\
FeI: & 6180 & SA12 \\
FeI & 6355 & SA12/H14 \\
FeI & 6394 & SA12/H14 \\
FeI & 6426 & SA12/H14 \\
CaI/FeII & 6451/6456 & H14/SA12 \\
CaI & 6494 & H14/SA12 \\
FeII & 6516 & H14/SA12 \\
H$\alpha$ & 6563 & SA12/SA15 \\
NiI/[NII]/CI & 6586/6583/6588 & SA12/H14/SA15 \\
FeII/HeI & 6679/6678 & H14/SA12/SA15 \\
\hline  
\end{tabular}
\tablefoot{The laboratory wavelengths are given. Uncertain identifications are marked by ":".
Lines that have been observed in EX Lupi during outburst \citep[SA12][]{sicilia12}
or quiescence \citep[SA15][]{sicilia15} and in the 2013 outburst of ASASSN-13db \citep[H14][]{holoien14} are marked in
the References column.}
\end{table}

\section{Accretion structure from spectroscopy\label{spectra-ana}}

\subsection{Line velocity auto- and cross-correlations}

The complexity of the emission lines requires a pixel-by-pixel analysis of the line structure,
which can be done by exploring the velocity-space via auto- and cross-correlations \citep{alencar01,sicilia15}.
The emitted flux at each velocity is cross-correlated with the flux at any other velocity, for either the
same line (auto-correlation), or for another line (cross-correlation).
This requires having numerous spectra with high S/N, which reduces our sample to
the very bright lines that are well-detected by both FEROS and CHIRON and that are not
blended with other lines within $\pm$300 km/s. By assigning each velocity bin to a gas parcel 
moving at a given rate, the connection between parts of the system is revealed by the auto- (and cross-) correlation matrix. 
This approach allows us to explore the lines in a non-parametric
way, which is particularly useful for very
complex lines such as H$\alpha$. The main limitation is that 
there can be several independent gas parcels moving at the same velocity with respect to our
line of sight (e.g., due to projection, or due to the presence of more than one accretion column).

We focused on H$\alpha$ and Fe II 4923 \AA\ as representative examples of the emission-only and strong inverse P-Cygni 
profiles. Lines from the same multiplet (e.g., Fe II 4923\AA\ and 5018 \AA) have nearly identical profiles 
and velocity structure \citep[as observed in EX Lupi][]{sicilia12}.
To construct the correlation matrices, the FEROS spectra were first resampled to the CHIRON resolution. For each line,
a local normalization was performed, measuring the continuum on both sides of the line in regions 
not affected by other features. We then obtained the auto-(cross-)correlation matrix by correlating each line pixel-by-pixel
with itself (a different line). We used customized Python routines based on 
the Spearman rank correlation to derive a correlation coefficient, r, and the false-alarm probability, p.

\begin{figure*}
\centering
\begin{tabular}{ccc}
\includegraphics[height=4.7cm]{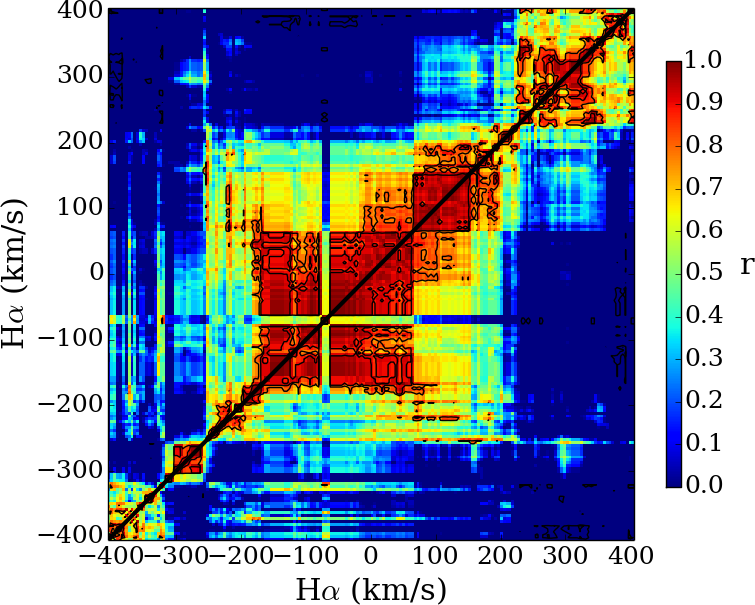} &
\includegraphics[height=4.7cm]{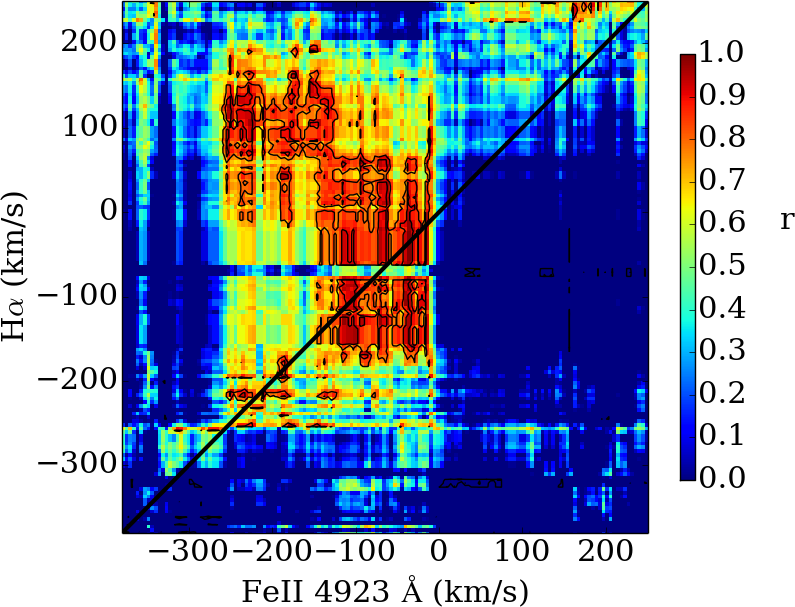} &
\includegraphics[height=4.7cm]{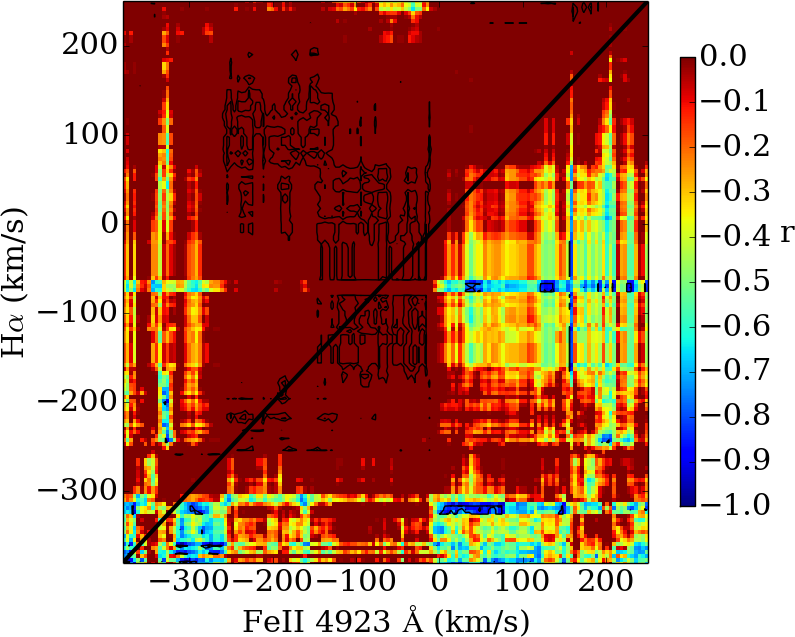} \\
\end{tabular}
\caption{Auto-correlation (left) and cross-correlation  of H$\alpha$ with the Fe II 4923\AA\ line; 
including positive (middle) and negative (right) correlations. 
The color scheme shows the value of the Spearman rank correlation parameter r, 
and the black contours mark the high-significance areas with false-alarm probability 1e-3 and 1e-4, respectively.}
\label{halphacorr-fig}%
\end{figure*}

Figure \ref{halphacorr-fig} shows the results of the correlations. The auto-correlation for the H$\alpha$ line 
reveals a complex asymmetry with a strong correlation between the 
blue- and red-shifted sides. The correlation coefficient between the line wings
and the center of the line decreases rapidly, suggesting that the zero-velocity gas components come from different 
or very extended locations. The cross-correlation matrix for H$\alpha$ and Fe II 4923\AA\ shows a 
correlation between the central and blue-shifted part of the line. There are no
evident correlations for line velocities between 0 and100 km/s, but the red-shifted side of H$\alpha$ is 
correlated with the blue-shifted side of Fe II 4923\AA. A mild anticorrelation between
the red-shifted wings in H$\alpha$ and the red-shifted Fe II absorption
suggests that the absorption becomes
deeper when the H$\alpha$ line wings are stronger. This is in agreement with an increase of the material along 
the line-of-sight, supporting the classification as a nearly edge-on YY Ori system. 
Being more extended and associated with lower temperatures, H$\alpha$  would increase at the same
time as the self-absorption in the Fe II lines increases.
There is no further evidence of relative correlations between the lines nor between the absorption and
wind components other than the general correlation that all lines tend to get stronger (or weaker) in parallel.


\subsection{Line velocity structure \label{linevel}}

To study the line structure, we need to estimate the radial velocity of ASASSN-13db. The complexity
of the emission line profiles and the lack of any reliable photospheric absorption line in the high-resolution spectra
do not allow for a direct determination, 
so we adopt as a reference the velocities of the 
L1615 and L1616 clouds in Orion \citep[22.3$\pm$4.6 km/s;][]{gandolfi08}, which are roughly similar to the average radial
velocities in the ONC \citep[$\sim$25-30 km/s;][]{sicilia05}. The lines show strong
day-to-day variability in both intensity and structure. 

Besides the inverse P-Cygni profile, the FEROS data from November 29, 2014 and December 4, 2014
reveal deep, blueshifted absorption features in several of the lines, including the H$\alpha$, Fe I, and Fe II lines
(see Figures \ref{halphahbeta-fig} and \ref{lineexamples-fig} and Table \ref{alllines-table}). 
The velocity of the blueshifted
absorption changes by about $\geq$50 km/s between the two dates, being about $\sim -$300 km/s on November 29, 2014
and between $\sim -$250/$-$200 km/s on December 4, 2014. Although there are small differences in the velocity from line
to line, the general behavior is consistent. Blueshifted absorption lines are
usually ascribed to winds. In this case, the rapid velocity change of the line at roughly
fixed strength is suggestive of rotational modulation in a non-axisymmetric wind. The
FEROS spectrum from November 17, 2014 does not have blueshifted absorption features below the continuum, although
it has a clear absorption feature that dominates the blue wing of the line (see Figure \ref{halphahbeta-fig}). 
The CHIRON spectra taken on November 30, 2014 (between the two FEROS spectra) have a marginal wind absorption
feature, but it is
hard to establish because of the low S/N. There is no apparent correlation between the wind and the rotation
phase, with the wind being observed at phase 0.51 and 0.73, but not at 0.65. 
This suggests a strong, variable wind in addition to the effects of rotation and geometry. More data 
would have been desirable to confirm the wind geometry.

We restrict the velocity and structure analysis to lines that are strong and unblended
over at least a $\pm$200 km/s region around the line, and that are well-identified
(i.e., we cannot reasonably attribute the same line to more than one species). 
We also restrict our analysis to the metallic lines, excluding 
H$\alpha$, which has a very complex profile. We concentrate on the high S/N FEROS data.

\begin{figure*}
\centering
\includegraphics[width=14.5cm]{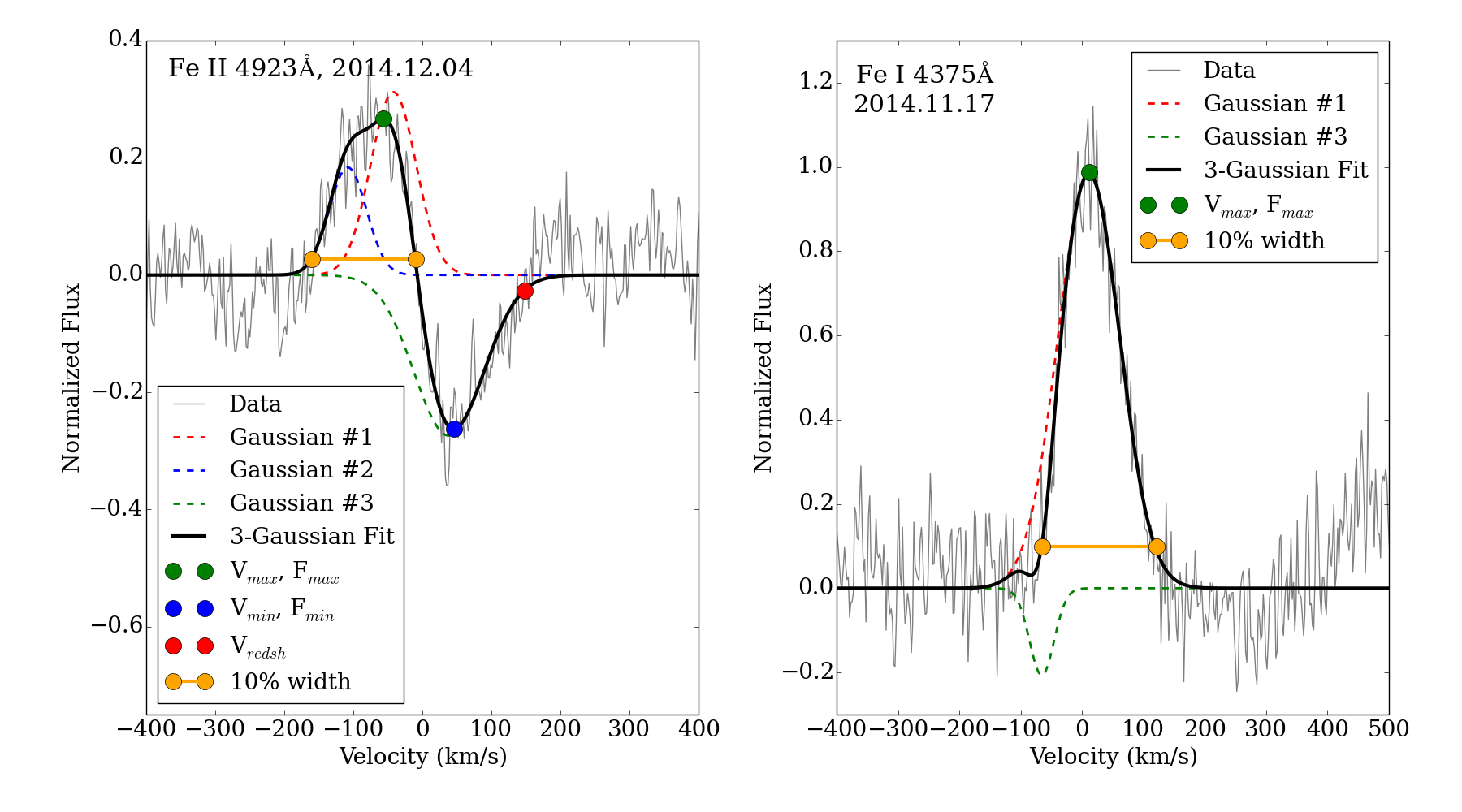}
\caption{Some examples of the three-Gaussian fits and the quantities derived from them. We note that
the Fe I 4375\AA\ line can be fitted with only two Gaussians, and does not include absorption component
parameters.}
\label{fitexamples-fig}
\end{figure*}

To perform a quantitative analysis of the line profile, we first fit the emission with a multi-Gaussian
profile \citep[as had been done for EX Lupi;][see also Appendix \ref{gaussian-app}]{sicilia12,sicilia15}.
The emission lines of ASASSN-13db are more complex than those of EX Lupi (which consist of well-defined broad and narrow
components), so
the fits are used to quantify the line in terms of the intensities and velocities of the emission
and absorption components, and the emission line width (see Figure \ref{fitexamples-fig}). Although the three-Gaussian components are
highly degenerate, the general line parameters derived are very robust,
and we do not observe any differences (within their errors) when derived from different three-Gaussian fits.
They include the normalized flux peak (F$_{max}$) and its
velocity (V$_{max}$), the velocity (V$_{min}$) and depth (F$_{min}$) of the redshifted absorption, 
the maximum velocity of the absorption feature (V$_{redsh}$; calculated as the maximum redshifted velocity at which the absorption 
correspond to 10\% of the maximum absorption), and the width at 10\% of the peak in the emission component (W$_{10\%}$). Some lines are 
well-fitted with only two Gaussians. For the lines
where the absorption component is not present, the line parameters related to absorption are not computed.
The errors in the derived quantities depend on the S/N of the spectrum and the line width. Errors in the maximum and minimum flux are
estimated based on the average S/N within the emission or absorption feature. Errors in the velocities are derived accordingly 
from the Gaussian fit, taking into account the peak and minimum flux errors. 
The uncertainties in W$_{10\%}$ and V$_{redsh}$ are derived considering the errors in the peak and minimum flux with respect to which they are
measured.

\begin{figure*}
\centering
\includegraphics[width=15cm]{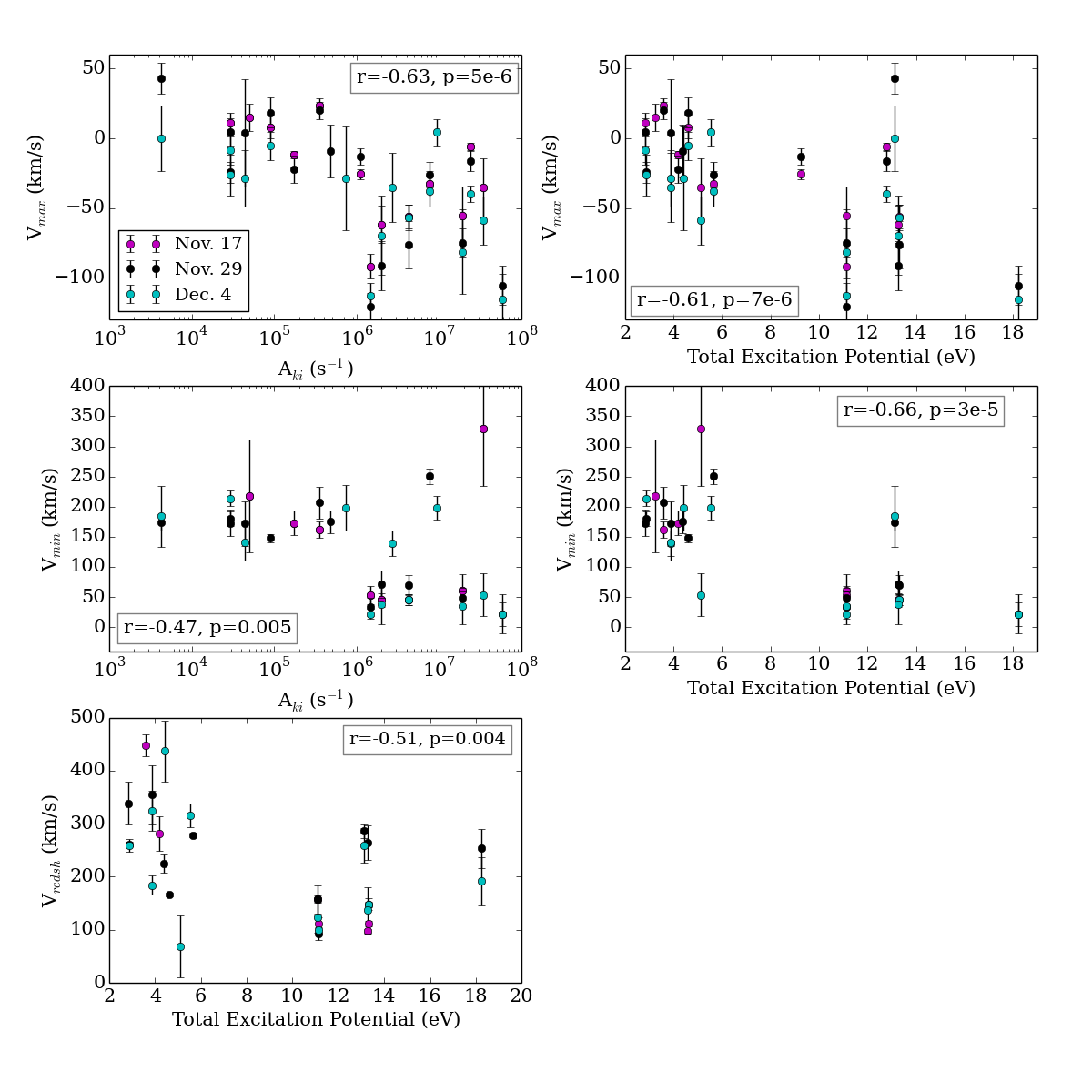}
\caption{Strong correlations between the line profiles and different atomic parameters. The correlation coefficient (r) and
the false-alarm probabilities (p) for the Spearman rank correlation are given in the plots. See text for
details. }
\label{correlation-fig}%
\end{figure*}

We used these quantifiers of the line shapes to explore potential 
correlations between the atomic parameters \citep[including the energies of the upper and lower levels, 
the transition probability, and the
sum of the ionization and excitation potentials;][]{bertout82} and the line emission and absorption properties. 
The strength of the correlation was determined using the Spearman rank test, which produces 
a correlation coefficient (r) and a false-alarm probability (p) for each pair of variables.
Negative correlation coefficients mean that the quantities are anticorrelated.
Among all non-trivial possibilities, five strong correlations (p$\leq$0.005; see Figure \ref{correlation-fig}) 
and six marginal correlations (p$\sim$0.006-0.06) arise, 
listed in Table \ref{correlation-table}.  
Considering the line parameters themselves, we find five further correlations. Three of them are trivial (e.g., all 
the velocities are correlated, indicating that lines tend to 
globally shift in velocity), while two marginal correlations
suggest that stronger lines tend to be more redshifted, and
that shallower absorption features tend to appear at larger redshifted velocities
(see Appendix \ref{correlations-appendix}).

\begin{table}
\caption{\label{correlation-table} Correlations and marginal correlations found between the
line properties and atomic parameters.   }
\centering
\begin{tabular}{lccl}
\hline\hline
Correlated quantities & r  & p   & Comments \\
 \hline
V$_{max}$ vs E$_k$  & $-$0.36 & 0.02 & marginal \\
V$_{max}$ vs A$_{ki}$ & $-$0.63 & 5e$-$6 & \\
V$_{max}$ vs Total Exc. & $-$0.61 & 7e$-$6 & \\
F$_{max}$ vs E$_i$ & $-$0.31 & 0.04 & marginal \\
F$_{max}$ vs E$_k$ & $-$0.39 & 0.01 & marginal \\
F$_{max}$ vs Total Exc. & $-$0.39 & 0.009 & marginal \\
V$_{min}$ vs E$_k$ & $-$0.34 & 0.06 & marginal \\
V$_{min}$ vs A$_{ki}$ & $-$0.47 & 0.005 & \\
V$_{min}$ vs Total Exc.& $-$0.66 & 3e$-$5 & \\
V$_{redsh}$ vs A$_{ki}$& $-$0.47 & 0.009 & marginal \\
V$_{redsh}$ vs Total Exc.& $-$0.51 & 0.004 &  \\
V$_{min}$ vs V$_{max}$ & 0.77 & 1e$-$7 & \\
V$_{max}$ vs V$_{redsh}$ & 0.68 & 3e$-$5 & \\
V$_{max}$ vs F$_{max}$ & 0.35 & 0.02 & marginal \\
V$_{min}$ vs V$_{redsh}$ & 0.70 & 1e$-$5 &  \\
V$_{min}$ vs F$_{min}$ & 0.31 & 0.08 & marginal\\
\hline  
\end{tabular}
\tablefoot{Correlation coefficient (r) and false-alarm probability (p)
from the Spearman rank test. Marginal correlations are labeled.}
\end{table}

We find anticorrelations between the transition probability (A$_{ki}$) and the 
line velocities (for the peak velocity, V$_{max}$,
the velocity at maximum absorption, V$_{min}$, and the zero redshifted velocity, V$_{redsh}$). 
The lines with higher transition probabilities
appear more blue-shifted. Similar anticorrelations arise between the total excitation potential 
(the sum of the ionization potential, which is zero for 
neutral lines, and the energy of the upper level) and the line velocities (V$_{max}$, V$_{min}$, 
V$_{redsh}$). The lines with lower
excitation potentials are more strongly redshifted. Surprisingly, this is opposite to the behavior 
found for other systems with YY-Ori-type line profiles,
such as SCrA and CoD-35$^o$10525, for which  \citet{bertout82} and \citet{petrov14} found that
lines formed at low temperatures are strongest at low infall velocity. In our case, 
the anticorrelation between the total
excitation potential and the velocity of maximum absorption rather indicates that the highest infall 
velocities are reached for the lines 
formed at lower temperatures. Optical depth and geometrical considerations
(such as more absorbing material in the more distant locations, which also would display lower infall velocities,
or occultation of the hottest and densest parts of the flow due to the inclination) could play
a role here. One solution to produce a density inversion would be to accumulate material
away from the star (for instance, at the edge of the magnetosphere).
Hot spots associated with magnetic reconnection, which have been suggested to explain high-energy variability
in eruptive stars \citep{hamaguchi12}, could also produce high density, high temperature, low velocity structures. 
In any case, more observations (including high-energy data, magnetic mapping, and high-resolution, 
high S/N spectroscopy) would be required to explore this possibility.

\begin{table*}
\footnotesize{
\caption{\label{lineparam-table} Line velocity structure and atomic parameters (from NIST) for the high S/N lines. Line absorption parameters are fit
only when the line has a redshifted absorption.}
\centering
\begin{tabular}{lccccccccccc}
\hline\hline
Species & E$_i$ & E$_k$ & E$_{ion}$ & A$_{ki}$ & V$_{max}$ & F$_{max}$ & V$_{min}$ & F$_{min}$ & W$_{10\%}$ & V$_{redsh}$ \\
        & (eV)  & (eV)  & (eV)      & (s$^{-1}$& (km/s)    &      & (km/s)    &       & (km/s)     & (km/s) \\
\hline
{\bf 2014.11.17} \\
FeI 4375.93 & 0.00 & 2.83 & 0.00 & 2.95E+04 & 11$\pm$8 & 0.989$\pm$0.011 & --- & --- & 186$\pm$4 & --- \\
TiII 4571.98 & 1.57 & 4.28 & 6.83 & 1.92E+07 & $-$56$\pm$21 & 0.339$\pm$0.062 & 61$\pm$27 & $-$0.259$\pm$0.056 & 128$\pm$23 & 157$\pm$26 \\
TiII 5129.15 & 1.89 & 4.31 & 6.83 & 1.46E+06 & $-$92$\pm$9 & 0.468$\pm$0.016 & 53$\pm$16 & $-$0.106$\pm$0.017 & 140$\pm$8 & 111$\pm$12 \\
FeI 4602.94 & 1.48 & 4.18 & 0.00 & 1.72E+05 & $-$12$\pm$3 & 0.514$\pm$0.006 & 173$\pm$21 & $-$0.034$\pm$0.007 & 228$\pm$4 & 282$\pm$33 \\
FeII 4923.92 & 2.89 & 5.41 & 7.90 & 4.28E+06 & $-$56$\pm$8 & 0.336$\pm$0.006 & 46$\pm$9 & $-$0.155$\pm$0.009 & 283$\pm$6 & 112$\pm$5 \\
FeII 5018.43 & 2.89 & 5.36 & 7.90 & 2.00E+06 & $-$62$\pm$13 & 0.348$\pm$0.012 & 45$\pm$11 & $-$0.222$\pm$0.018 & 222$\pm$15 & 98$\pm$3 \\
MgI 5172.68 & 2.71 & 5.11 & 0.00 & 3.37E+07 & $-$35$\pm$21 & 0.375$\pm$0.032 & 329$\pm$96 & $-$0.110$\pm$0.018 & 204$\pm$28 & 669$\pm$129 \\
FeI 5506.78 & 0.99 & 3.24 & 0.00 & 5.01E+04 & 15$\pm$10 & 0.440$\pm$0.018 & 217$\pm$94 & $-$0.038$\pm$0.010 & 138$\pm$9 & 538$\pm$154 \\
FeI 6200.31 & 2.61 & 4.61 & 0.00 & 9.06E+04 & 8$\pm$8 & 0.522$\pm$0.008 & --- & --- & 201$\pm$7 & --- \\
FeI 6336.82 & 3.87 & 5.64 & 0.00 & 7.71E+06 & $-$33$\pm$9 & 0.512$\pm$0.012 & --- & --- & 179$\pm$9 & --- \\
FeI 6678.88 & 10.93 & 12.79 & 0.00 & 2.40E+07 & $-$6$\pm$3 & 0.680$\pm$0.003 & --- & --- & 221$\pm$2 & --- \\
CaII 8498.02 & 1.69 & 3.15 & 6.11 & 1.11E+06 & $-$26$\pm$3 & 1.642$\pm$0.007 & --- & --- & 297$\pm$5 & --- \\
FeI 8824.22 & 2.19 & 3.60 & 0.00 & 3.53E+05 & 24$\pm$5 & 0.733$\pm$0.007 & 162$\pm$14 & $-$0.233$\pm$0.004 & 135$\pm$2 & 448$\pm$21 \\
{\bf 2014.11.29} \\
FeI 4375.93 & 0.00 & 2.83 & 0.00 & 2.95E+04 & 5$\pm$10 & 0.431$\pm$0.006 & 173$\pm$21 & $-$0.068$\pm$0.008 & 338$\pm$11 & 339$\pm$40 \\
FeI 4461.65 & 0.08 & 2.87 & 0.00 & 2.95E+04 & $-$24$\pm$8 & 0.387$\pm$0.005 & 180$\pm$12 & $-$0.167$\pm$0.007 & 321$\pm$19 & 261$\pm$4 \\
TiII 4571.98 & 1.57 & 4.28 & 6.83 & 1.92E+07 & $-$75$\pm$10 & 0.276$\pm$0.010 & 48$\pm$12 & $-$0.283$\pm$0.009 & 144$\pm$7 & 158$\pm$7 \\
TiII 5129.15 & 1.89 & 4.31 & 6.83 & 1.46E+06 & $-$121$\pm$10 & 0.275$\pm$0.010 & 34$\pm$15 & $-$0.080$\pm$0.012 & 155$\pm$6 & 92$\pm$11 \\
FeI 4602.94 & 1.48 & 4.18 & 0.00 & 1.72E+05 & $-$22$\pm$10 & 0.313$\pm$0.006 & 900$\pm$42 & $-$0.047$\pm$0.001 & 235$\pm$6 & --- \\
FeII 4923.92 & 2.89 & 5.41 & 7.90 & 4.28E+06 & $-$76$\pm$17 & 0.336$\pm$0.011 & 70$\pm$17 & $-$0.130$\pm$0.015 & 277$\pm$18 & 148$\pm$11 \\
FeII 5018.43 & 2.89 & 5.36 & 7.90 & 2.00E+06 & $-$91$\pm$18 & 0.265$\pm$0.016 & 72$\pm$23 & $-$0.186$\pm$0.013 & 189$\pm$15 & 264$\pm$32 \\
FeI 5332.90 & 1.55 & 3.88 & 0.00 & 4.36E+04 & 4$\pm$38 & 0.098$\pm$0.032 & 172$\pm$37 & $-$0.165$\pm$0.026 & 184$\pm$63 & 355$\pm$56 \\
FeII 5991.38 & 3.15 & 5.22 & 7.90 & 4.20E+03 & 43$\pm$11 & 0.172$\pm$0.005 & 174$\pm$13 & $-$0.093$\pm$0.005 & 174$\pm$7 & 286$\pm$13 \\
FeI 6200.31 & 2.61 & 4.61 & 0.00 & 9.06E+04 & 18$\pm$11 & 0.267$\pm$0.006 & 148$\pm$7 & $-$0.049$\pm$0.013 & 212$\pm$9 & 167$\pm$4 \\
FeI 6336.82 & 3.87 & 5.64 & 0.00 & 7.71E+06 & $-$26$\pm$9 & 0.426$\pm$0.004 & 251$\pm$13 & $-$0.031$\pm$0.005 & 795$\pm$20 & 279$\pm$5 \\
SiII 6347.10 & 8.12 & 10.07 & 8.15 & 5.84E+07 & $-$105$\pm$14 & 0.178$\pm$0.015 & 22$\pm$20 & $-$0.165$\pm$0.009 & 110$\pm$7 & 253$\pm$37 \\
FeI 6393.60 & 2.43 & 4.37 & 0.00 & 4.81E+05 & $-$9$\pm$19 & 0.337$\pm$0.029 & 175$\pm$19 & $-$0.122$\pm$0.035 & 182$\pm$21 & 224$\pm$17 \\
FeI 6678.88 & 10.93 & 12.79 & 0.00 & 2.40E+07 & $-$16$\pm$8 & 0.369$\pm$0.001 & --- & --- & 294$\pm$4 & --- \\
CaII 8498.02 & 1.69 & 3.15 & 6.11 & 1.11E+06 & $-$13$\pm$6 & 1.335$\pm$0.006 & --- & --- & 353$\pm$5 & --- \\
FeI 8824.22 & 2.19 & 3.60 & 0.00 & 3.53E+05 & 20$\pm$6 & 0.533$\pm$0.007 & 207$\pm$27 & $-$0.185$\pm$0.004 & 144$\pm$3 & 525$\pm$28 \\
{\bf 2014.12.04} \\
FeI 4461.65 & 0.08 & 2.87 & 0.00 & 2.95E+04 & $-$26$\pm$15 & 0.402$\pm$0.037 & 214$\pm$13 & $-$0.180$\pm$0.033 & 176$\pm$20 & 258$\pm$12 \\
FeI 4375.93 & 0.00 & 2.83 & 0.00 & 2.95E+04 & $-$8$\pm$10 & 0.401$\pm$0.014 & --- & --- & 384$\pm$24 & --- \\
TiII 4571.98 & 1.57 & 4.28 & 6.83 & 1.92E+07 & $-$81$\pm$30 & 0.282$\pm$0.096 & 35$\pm$30 & $-$0.314$\pm$0.091 & 135$\pm$40 & 123$\pm$29 \\
FeII 4923.92 & 2.89 & 5.41 & 7.90 & 4.28E+06 & $-$57$\pm$9 & 0.266$\pm$0.007 & 45$\pm$9 & $-$0.262$\pm$0.007 & 151$\pm$4 & 148$\pm$5 \\
FeII 5018.43 & 2.89 & 5.36 & 7.90 & 2.00E+06 & $-$70$\pm$28 & 0.285$\pm$0.088 & 38$\pm$33 & $-$0.229$\pm$0.091 & 164$\pm$30 & 137$\pm$44 \\
TiII 5129.15 & 1.89 & 4.31 & 6.83 & 1.46E+06 & $-$113$\pm$9 & 0.251$\pm$0.014 & 22$\pm$8 & $-$0.117$\pm$0.012 & 115$\pm$10 & 99$\pm$6 \\
MgI 5167.32 & 1.48 & 3.88 & 0.00 & 2.72E+06 & $-$35$\pm$25 & 0.584$\pm$0.043 & 139$\pm$21 & $-$0.155$\pm$0.073 & 264$\pm$34 & 184$\pm$18 \\
MgI 5172.68 & 2.71 & 5.11 & 0.00 & 3.37E+07 & $-$59$\pm$17 & 0.201$\pm$0.018 & 54$\pm$35 & $-$0.005$\pm$0.036 & 173$\pm$25 & 69$\pm$58 \\
FeI 5332.90 & 1.55 & 3.88 & 0.00 & 4.36E+04 & $-$29$\pm$20 & 0.128$\pm$0.011 & 141$\pm$31 & $-$0.104$\pm$0.011 & 226$\pm$67 & 325$\pm$37 \\
FeII 5991.38 & 3.15 & 5.22 & 7.90 & 4.20E+03 & 0$\pm$23 & 0.079$\pm$0.023 & 184$\pm$50 & $-$0.078$\pm$0.037 & 564$\pm$513 & 259$\pm$33 \\
FeI 6191.56 & 2.43 & 4.43 & 0.00 & 7.41E+05 & $-$29$\pm$37 & 0.102$\pm$0.017 & 198$\pm$38 & $-$0.139$\pm$0.013 & 219$\pm$38 & 437$\pm$58 \\
FeI 6200.31 & 2.61 & 4.61 & 0.00 & 9.06E+04 & $-$6$\pm$10 & 0.183$\pm$0.004 & --- & --- & 205$\pm$7 & --- \\
FeI 6336.82 & 3.87 & 5.64 & 0.00 & 7.71E+06 & $-$38$\pm$11 & 0.263$\pm$0.004 & --- & --- & 382$\pm$17 & --- \\
SiII 6347.10 & 8.12 & 10.07 & 8.15 & 5.84E+07 & $-$116$\pm$19 & 0.124$\pm$0.016 & 22$\pm$32 & $-$0.085$\pm$0.012 & 140$\pm$22 & 192$\pm$45 \\
FeI 6400.00 & 3.60 & 5.54 & 0.00 & 9.27E+06 & 4$\pm$10 & 0.393$\pm$0.010 & 198$\pm$20 & $-$0.102$\pm$0.009 & 171$\pm$8 & 316$\pm$22 \\
FeI 6678.88 & 10.93 & 12.79 & 0.00 & 2.40E+07 & $-$40$\pm$6 & 0.237$\pm$0.002 & --- & --- & 323$\pm$5 & --- \\
\hline  
\end{tabular}
}
\end{table*}

\subsection{Accretion column properties derived from the emission lines \label{physicalproperties-sect}}

For stars with a large number of emission lines, it is possible to determine the physical conditions
(temperature and density) in the accretion columns by using line ratios of neutral and ionized lines \citep{beristain98,sicilia12, sicilia15}.
Although the S/N for most of the data is too low to perform a velocity-dependent analysis,
the relative intensities can be explored for a number of lines.
As with EX Lupi, we can constrain the approximate density in the accretion column based on
the observed typical accretion rate \.{M}=2$\times$10$^{-7}$M$_\odot$/yr. 
This accretion rate would result from the mass within the volume of the accretion column multiplied by the approximate
density. The volume that is accreted onto the star per second 
can be approximated by the typical velocity of the infalling material ($v\sim40$ km/s) multiplied by the cross-section of the column(s). Therefore, the accretion rate can be written as
\begin{equation}
\dot{M}= \mu \, n  \, v \, f 4 \pi R_*^2, \label{eq:mdot}
\end{equation}
where $n$ is the density, $\mu$ is the mean atomic weight, $f$ is the fraction of the stellar surface
covered by spots \citep[in general, a small part of the stellar surface, 1-20\%][]{calvet98,lima10}, and R$_*$ is the stellar radius
\citep[1.1$R_\odot$ as estimated by][]{holoien14}. These values imply a particle 
(mostly H) density in the range 
n$\sim$2$\times$10$^{13}$-4$\times$10$^{14}$ cm$^{-3}$. If we assume that, at the relevant temperatures in
the accretion column, hydrogen and helium will be mostly neutral and all the metals will be
singly ionized, then the electron density will be a factor of 1000 lower,
n$_e \sim$2$\times$10$^{10}$ to 4$\times$10$^{11}$ cm$^{-3}$. 

An independent constraint 
on the electron density can be obtained from the saturation of the Ca II IR
triplet lines \citep{hamann92}. For the Ca II IR lines to have similar strength, collisional decay
(given by C$_{ki}$)
must dominate over the radiative transition rate (A$_{ki}$), which requires n$_e$C$_{ki} \gg$A$_{ki}$/$\tau$.
This imposes a relation between the opacity and the electron density of $\tau$\,n$_e \gtrsim$ 10$^{13}$ cm$^{-3}$ for Ca II 8542/8662\AA. 
Considering that usually $\tau >$1-10 \citep{grinin88,shine74}\footnote{In our case, the column density is
larger (although likely not by much) than 1 to allow for the self-absorption.},
we arrive to an approximate value for  n$_e \sim$ 10$^{12}$ cm$^{-3}$.
These values are slightly higher than those derived from the accretion rate. 
In the case of EX Lupi, the mismatch between
accretion-based estimates and the requirements for Ca II saturation was attributed to
higher levels of ionization than expected due to UV radiation from the accretion shock.
For ASASSN-13db, we expect less ionizing radiation due to the lower mass of the system,
and both values agree within a factor of few. The accretion columns could also be
significantly optically thicker for the case of a small object viewed nearly
edge-on, so the differences are reasonable given the uncertainties.

Several lines can be used to constrain the temperature and density.
If we assume local thermodynamical equilibrium, the relative populations of ions and neutrals is determined 
by the Saha equation \citep{saha21,mihalas78}
 \begin{equation}
        \frac{N_{j+1}n_e}{N_j} 
        = 
        \left(\frac{2\pi m k T}{h^2}\right)^{3/2}
        \frac{2 U_{j+1}(T)}{U_j(T)}
        e^{-\chi_I / kT}.
\label{eq:saha}
\end{equation}
Here, N$_{j+1}$, N$_j$, and n$_e$ represent the number of atoms in the j+1 and j
ionization states and the electron number density, T is the temperature, m is the electron
mass, $\chi_I$ is the ionization potential, and U$_{j+1}$ and U$_j$ are the partition functions for the
j+1 and j states.  The level populations follow a
Boltzmann distribution, and can be transformed into
line intensity ratios to compare with the observed data.
Using the density threshold imposed by the Ca II emission, we used the Saha equation 
\citep[together with the data available from NIST;][]{ralchenko10} for several line pairs to
further constrain the temperature and density.

The lack of He I emission\footnote{The only line that could be attributed to He I
is at 6678\AA. Since there is no He I emission at 5875\AA, although
both lines have very similar transition probabilities and a common upper level (so the
5875\AA\ line would be expected to be a factor of few stronger than the line at 6678\AA), we
conclude that the emission at 6678\AA\ is most likely due to Fe II.} also constrains the temperatures within the
accretion flow. The He I line at 5875\AA\ is commonly observed in accreting, low-mass stars
\citep{hamann92,sicilia05}. Temperatures around 20000 K are required to thermally excite the He I line,
which suggests that the accretion shock/flow in ASASSN-13db is cooler than in solar-type objects.

Ti I/Ti II lines are another strong indicator of temperature. For a temperature of 6500 K, Ti I emission
disappears unless the density is high (n$_e \sim$ 1$\times$10$^{14}$ cm$^{-3}$). 
A temperature of 5000 K would require a density of around n$_e \sim$ 3$\times$10$^{11}$ cm$^{-3}$, while
for 5800 K, the expected density for substantial Ti I emission would be
n$_e \sim$ 1$\times$10$^{13}$ cm$^{-3}$. From the constraints derived from the accretion
rate and the Ca II emission, 
we conclude that the temperature in the accretion structures is most likely below 6000 K.

The observed 1:2 Fe I 5269\AA/Fe II 5169\AA\ line ratio is also consistent with a temperature 
around 5800 K for a density in the range n$_e \sim$5$\times$10$^{11}$-2$\times$10$^{12}$ cm$^{-3}$. 
These low temperatures in the accretion flow 
are consistent with the M5 spectral type of
the object \citep{holoien14}, which for young stars lies close to the border between
very-low-mass stars and brown dwarfs.

The lack of H$\beta$ emission is in agreement with this picture; according
to the accretion flow models by \citet{muzerolle01}, absorption in H$\beta$ appears at
large inclinations ($\gtrsim$75 degrees)
for a flow temperature of about 6000 K and an accretion rate around 10$^{-8}$M$_\odot$/yr
(lower than we observe). Although the temperature is in agreement with our estimates,
our accretion rate during the 2014-2017 outburst is substantially higher. 

All these lines of evidence suggest that the
dominant temperature in the flow is of the order of 5800-6000 K. 
Detailed line radiative transfer models, applied to this specific case of a very-low-mass object, should be
explored in the future to address the line structure.


\section{Discussion: the variable nature of ASASSN-13db \label{discussion}}

The 2013 and 2014-2017 outbursts of ASASSN-13db are significantly different.
While the first outburst is in full agreement with typical EXor
outbursts \citep[e.g.,][]{herbig01},
the second one is unusual in length and in magnitude. 
The length of the 2014-2017 outburst is shorter than typical FUors,
although several FUor outbursts lasting only a few years have been observed 
\citep[e.g., ZCMa, V1647 Ori, $\lbrack$CTF93$\rbrack$216-2;][]{fedele07,aspin11,caratti11,audard14,bonnefoy16}. 
Objects with similar outburst lengths are often referred to as EXors, but this is typically accompanied by a question mark
or considered as an intermediate case,
especially if some FUor characteristics are present \citep[e.g.,][]{abraham04,fedele07,caratti11}.
Although ASASSN-13db experienced a slow decay after September 2016 \citep[similar to the smooth,
exponential-like decays observed in FUors;][]{hartmann96}, the
final abrupt dimming and return to quiescence over $\sim$2 months is more typical of EXor behavior. 
The overall shape of the light curve (Figure \ref{ASAScurve-fig}) strongly resembles, in length, duration, and
general shape (including the quick magnitude drop after a slow fading), that of the light curve of 
the FUor variable V1647 Ori \citep{fedele07}, the object that illuminated the McNeil nebula during its 2004 
outburst \citep{mcneil04,briceno04}. The 2010 outburst of the very-low-mass star [CTF93]216-2 \citep[M$_*$=0.25 M$_\odot$;][]{caratti11}
also shares many characteristics with the ASASSN-13db 2014-17 outburst.
An intermediate class, MNors, has been proposed for objects similar to the
McNeil nebula \citep{contreras17}, although for ASASSN-13db there is no evidence for a reflection nebula
(see Figure \ref{LCOGTrgb-fig}), and the object appears less embedded than other MNors.
The accretion rate observed in outburst ($\sim$2$\times$10$^{-7}$M$_\odot$/yr) and the
increase of accretion between quiescence and outburst of at least two orders of magnitude
are more in agreement with large EXor outbursts, such as the 2008 EX Lupi outburst \citep{sicilia12,juhasz12}.
If the second outburst corresponds to a FUor episode, it would be the first time that
an object has been seen to undergo both types of behavior. ASASSN-13db would also be the
lowest mass FUor object known to date, suggesting that accretion outbursts, like the rest of disk and accretion
properties, occur in very-low-mass stars and perhaps substellar-mass objects.

Few-day periodic or quasi-periodic signatures, 
usually attributed to inner disk structures, 
are characteristic of FUor objects \citep[e.g.,][]{herbig03,siwak13}. 
For the periodic 4.15d signature to arise in the inner disk, we would need a
disk structure located at about 0.09 au or about 20 R$_\odot$, assuming Keplerian rotation. 
Assuming T$_{eff}$=3075 K for a M5 star, with a radius of about 1.1 R$_\odot$ \citep{holoien14}, the temperature in this
region would be of the order of $\sim$730 K (strongly dependent on dust properties and inner disk structure), which
is fully compatible with the presence of silicate dust grains. However, if the typical temperature in outburst
is higher ($>$4800 K),
the disk temperature at 0.09 au would be at least 1100-1300 K, close to the
silicate sublimation temperature of $\sim$1500 K. 
ASASSN-13db may be near edge-on or at a high angle, so extinction by accretion channels or 
extended accretion structures in the disk cannot be excluded.
Nevertheless, the relatively smooth, colorless sinusoidal light curve 
is rather suggestive of modulations induced by accretion-related hot/cold spot(s), rather than periodic eclipses. 

The detection of the 4.15 d period in data taken during a period of over 500 days (epochs B and C)
suggests that the mechanism responsible for the variability is very stable, or has
only experienced small variations during this time. A stable accretion structure has been
observed in V1467 Ori \citep{hamaguchi12} and EX Lupi \citep{sicilia15}. Very-low-mass objects are
expected to have strong and mostly bipolar magnetic fields \citep{morin10,gregory14}, which could produce
these structures.
The modulation becomes undetectable in January 2017, once the light curve is
dominated by a rapid decrease in magnitude, but appears to be again consistent with the small-scale variability
observed in March 2017 (see Figure \ref{phasefoldedLCOGT-fig}, bottom).

The variations in line depth and wind absorption suggest that both the accretion structures and the associated
wind component are not uniformly distributed around the star. 
A non-axisymmetric location could explain the rapid,
day-to-day variations in the velocity of the
wind and redshifted absorption component due to rotation. Although rotational 
modulations induced by non-axisymmetric accretion columns
are common \citep[e.g.,][]{costigan12,kurosawa13}, wind modulations are rare. If associated with the accretion column,
the observations of ASASSN-13db could be consistent with the X-wind scenario \citep{shu94} or a jet
originating at the magnetic reconnection point proposed to explain the high-energy variability of V1647 Ori 
\citep{hamaguchi12}. Further data is required to explore this possibility.

Despite the  potentially high inclination of the disk (or, at least, a view of the accretion columns along the infall direction), 
several of the observations rule out the possibility that
the observed behavior is caused by eclipses or occultations by disk material or UXor-type variability \citep[e.g.,][]{grinin91,natta99}.
UXor variability can produce 
apparently "high" states over long periods of time \citep{bouvier13}, but UXors present high extinction and trace a ">" shaped
curve in the color-magnitude diagram, since the colors become bluer at minimum 
due to scattering. The observed changes in color, veiling, and emission lines, 
and the fact that the photospheric spectrum of the star is only seen in quiescence and corresponds to a M5 
pre-main sequence star with age-appropriate luminosity and radius (see Section \ref{mdot-sect}),
are strongly suggestive of accretion eruptions. Moreover, the characteristics of the 
2013 outburst and the low extinction of the source \citep{holoien14}
rule out UXor variability as an explanation of the light curve.

\begin{table}
\footnotesize{
\caption{\label{dip-table} Magnitude dips observed in the ASASSN light curve. }
\centering
\begin{tabular}{lcccc}
\hline\hline
JD  & V$_{obs}$  & V$_{median}\pm\sigma_{std}$  & Depth & Epoch \\
(d) & (mag) & (mag) & (mag)& \\
\hline
2457103.532 & 13.92$\pm$0.03 & 13.39$\pm$0.25 & 0.54 & A\\
2457103.533 & 13.91$\pm$0.03 & 13.39$\pm$0.25 & 0.52 & A \\
2457367.613 & 14.10$\pm$0.03 & 13.59$\pm$0.16 & 0.51 & B \\
2457706.855 & 14.94$\pm$0.16 & 14.34$\pm$0.39 & 0.59 & C \\
\hline  
\end{tabular}
\tablefoot{Dips are defined against the
median value within $\pm$5 days, given in the table as V$_{median}$, together with the standard deviation ($\sigma_{std}$).
The letters A, B, and C indicate the epoch when the dip was observed. }
}
\end{table}

Nevertheless, a couple of rapid dimmings or dips, usually lasting 1-2 days,
are also observed, during which the luminosity drops by about 0.5 mag.
The dip on JD 2457706 is also detected in $i'$ in the Kent data from 2457707. 
One potential small dip, ranging between 0.7 and 0.9 mag, may also be observable in the $g'$ band (but not in $r'$ nor $i'$)
during the post-outburst phase (see Figure \ref{phasefolded-fig}), although since the $g'$ magnitude is
still decreasing and the color is changing at the time of the observations it is hard to establish whether it is a real
occultation event.
If the dips are due to extinction by circumstellar material, it is possible to make a rough estimate of the 
mass required for the events. Assuming that the disk feature that produces the events is located at the
inner disk rim at the dust sublimation radius ($\sim$0.077 au for an outburst temperature of $\sim$5800K),
that it covers a significant part of the orbit,
and that the typical increase in extinction is A$_V$= 0.5\,mag,
\citep[which corresponds to a column density of $\sim$1.0$\times$10$^{21}$ cm$^{-2}$;][]{bohlin78}, 
a total mass of $\sim$10$^{-7}$ M$_\oplus$ needed can be estimated by the product of the volume of the structure 
times the density (where the density
would be roughly the column density divided by the depth of the structure).
The orbital timescale at the dust sublimation radius is $\sim$20 d. Since the small eclipses or dips are not periodic, 
the material must be undergoing rapid variations from orbit to orbit.


\section{Summary and conclusions \label{conclu}}

ASASSN-13db shares characteristics with several types of young variable star, experiencing
frequent, strong outbursts with variable temporal scales. 
Our results are summarized below:
\begin{itemize}
\item Since its discovery in 2013, ASASSN-13db has experienced two distinct outbursts with very different timescales: while the
2013 outburst appears to be a typical EXor accretion episode, the length of the 2014-2017 outburst is extreme 
by the standard of EXors and intermediate between EXor and FUor episodes. 
ASASSN-13db adds to the growing evidence that not all 
eruptive accreting stars can be easily ascribed to one of the two types. 
\item The emission line spectrum during the 2014-2017 outburst is very similar to the outburst spectrum of EX Lupi in
2008 \citep{sicilia12}, and shows day-to-day variability. 
The post-outburst spectrum is very similar to the 2013 outburst spectrum \citep{holoien14}, and includes
many lines that have also been observed in EX Lupi in both outburst and quiescence \citep{sicilia15}.
\item The prominent inverse P-Cygni profiles displayed by the emission lines suggest that the system is nearly edge-on.
\item The light curve of ASASSN-13db during outburst is modulated with a 4.15d period, probably
due to stellar rotation. The presence of a rotational modulation during a time when the object is
on a high accretion state suggests that the accretion column(s) are few in number and relatively stable.
A similar situation was observed for EX Lupi \citep{sicilia15}, although the number or
structure of the accretion structures may have changed between epochs A and B in the case of ASASSN-13db.
\item A strong wind absorption component is observed in two of the FEROS spectra. The observed 
rapid velocity changes 
could indicate a non-axisymmetric wind, probably related to accretion variations and rotationally modulated. 
A wind that is strongly coupled to the accretion column would 
also be highly non-axisymmetric, as in the X-wind scenario \citep{shu94} or
wind/jet associated with magnetic reconnection \citep{hamaguchi12}.
\item The presence of a strong wind component that produces a blue-shifted absorption feature below the continuum
also suggests that the spectral features observed during the 2014-2017 episode are intermediate between EXor and FUor
outbursts. The extreme inclination of the system is likely to play a role in the observed profiles, especially
regarding the red-shifted absorption and may prevent the non-axisymmetric wind from being visible at
all times.
\item Analyzing the relative intensities of the neutral and ionized lines, we find that the accretion flow 
is cooler than
in EX Lupi or in typical CTTS models, with a typical temperature in the range of 5800-6000 K.
Since ASASSN-13db is the EXor with the latest spectral type known to date,
this would suggest a downscaling of temperatures and accretion flow densities with the stellar mass.
\item The high variability observed since its discovery, together with the particularly low mass of ASASSN-13db, make
it an ideal case to further explore accretion outbursts in young stars.
\end{itemize}

\begin{acknowledgements}
A.O. acknowledges support by the Royal Astronomical Society via a 2016 Summer Fellowship
under the supervision of A.S.A.
Support for J.L.P. is provided in part by FONDECYT through the grant 1151445 and by the 
Ministry of Economy, Development, and Tourism's Millennium Science Initiative through grant IC120009, 
awarded to The Millennium Institute of Astrophysics, MAS. 
T.W.-S.H. is supported by the DOE Computational Science Graduate Fellowship, grant number DE-FG02-97ER25308.
B.J.S. is supported by NASA through Hubble Fellowship grant HST-HF-51348.001 awarded by the Space Telescope 
Science Institute, which is operated by the Association of Universities for Research in Astronomy, Inc., for NASA, under contract NAS 5-26555.\\

We are very grateful to K. Rowlands for her help with A.O. fellowship application, and to
 A. Jord\'{a}n for providing the CHIRON spectral observations.
We are very grateful to the amateur astronomers who have provided
data for the project, Roger Pickard (The British Astronomical Association, Burlington House, 
Piccadilly, London W1J 0DU) and Georg Piehler (Selztal Observatory, Friesenheim, Bechtolsheimer Weg, Germany 
and Physikalischer Verein, Gesellschaft für Bildung und Wissenschaft, 
Frankfurt am Main, Robert-Mayer-Str. 2-4, Germany). We also thank the anonymous 
referee for his/her review of the manuscript, and Annelies Mortier 
and Silvia Alencar for useful discussion and suggestions.\\

This work makes use of observations from the LCO network.
ASAS-SN is supported by the Gordon and Betty Moore Foundation through 
grant GBMF5490 to the Ohio State University and NSF grant AST-1515927. We 
thank Las Cumbres Observatory and its staff for their continued support of 
ASAS-SN. Development of ASAS-SN has been supported by NSF grant 
AST-0908816, the Center for Cosmology and AstroParticle Physics at the 
Ohio State University, the Mt. Cuba Astronomical Foundation, the Chinese 
Academy of Sciences South America Center for Astronomy (CASSACA), and by 
George Skestos. 
This research was made possible through the use of the AAVSO Photometric All Sky Survey (APASS) 
funded by the Robert Martin Ayers Sciences Fund, data provided by Astrometry.net \citep{barron08}, 
and filter curves from the Visual Observatory.
This research makes use of SDSS data. Funding for the SDSS and SDSS-II has been provided by the Alfred P. Sloan Foundation, the Participating Institutions, the National Science Foundation, the U.S. Department of Energy, the National Aeronautics and Space Administration, the Japanese Monbukagakusho, the Max Planck Society, and the Higher Education Funding Council for England. The SDSS Web Site is http://www.sdss.org/.
The SDSS is managed by the Astrophysical Research Consortium for the Participating Institutions. The Participating Institutions are the American Museum of Natural History, Astrophysical Institute Potsdam, University of Basel, University of Cambridge, Case Western Reserve University, University of Chicago, Drexel University, Fermilab, the Institute for Advanced Study, the Japan Participation Group, Johns Hopkins University, the Joint Institute for Nuclear Astrophysics, the Kavli Institute for Particle Astrophysics and Cosmology, the Korean Scientist Group, the Chinese Academy of Sciences (LAMOST), Los Alamos National Laboratory, the Max-Planck-Institute for Astronomy (MPIA), the Max-Planck-Institute for Astrophysics (MPA), New Mexico State University, Ohio State University, University of Pittsburgh, University of Portsmouth, Princeton University, the United States Naval Observatory, and the University of Washington. 
\end{acknowledgements}







\Online

\onecolumn

\begin{appendix}

\section{Fitting and extracting the line velocity parameters \label{gaussian-app}}

In this section, we describe three-Gaussian model fits performed on the strong emission lines in order to
analyze and quantify their velocities. Only lines that were not contaminated by nearby features, atmospheric
absorption, or bad pixels were modeled. This means that some of the lines are excluded on certain dates.
The lines were first normalized using the
feature-free continuum on both sides of the line, obtaining the normalized flux ($F_{norm}$) for each velocity. Any bad pixels were excluded from the fit, 
using an interactive Python routine. Finally, the lines were fit with a three-Gaussian model where the fitted
normalized flux ($F_{fit}$) is written as a function of the line velocity, $v$,
\begin{equation}
F_{fit}(v) = \sum_{n=1}^{n=3} A_n e^{(v-v_n)^2/\sigma_n^2}. \label{gauss-formula}
\end{equation}
Here, $A_n$ is the amplitude of the Gaussian, $v_n$ is the zero-point velocity, and $\sigma_n$ is the
Gaussian width. Since most of the lines have an absorption component, we force the third Gaussian to be in absorption
($A_3<0$). Depending on the line profile and S/N, some of the lines are fit with only two Gaussian 
components. Although the fits do not have a physical interpretation and are strongly degenerate, they allow us to quantify
the observed velocities and to explore the relations between velocities and atomic parameters (see Section \ref{linevel}).
Table \ref{gaussians-table} contains the fit values for each one of the 
lines.

\begin{longtable}{lccccccccc}
\caption{Gaussian fits to the selected strong lines observed in the FEROS data. Parameters as in Equation \ref{gauss-formula}.
 \label{gaussians-table}} \\
\hline
\hline 
Species/$\lambda$ &  $A_1$ & $\sigma_1$  & $v_1$ &  $A_2$ & $\sigma_2$  & $v_2$ &  $A_3$ & $\sigma_3$ & $v_3$ \\
         (\AA)     &        & (km/s) & (km/s)    &        & (km/s) & (km/s)    &        & (km/s) & (km/s)    \\
\hline
\endfirsthead
\caption{Continued.} \\
\hline
\hline
Species/$\lambda$ &  $A_1$ & $\sigma_1$  & $v_1$ &  $A_2$ & $\sigma_2$  & $v_2$ &  $A_3$ & $\sigma_3$ & $v_3$  \\
         (\AA)     &        & (km/s) & (km/s)    &        & (km/s) & (km/s)    &        & (km/s) & (km/s)   \\

\hline
\endhead
{\bf 2014-11-17}\\
FeI 4375.93 & 0.985 &  71.209 &  11.061 &  0.299 &  419.829 &  $-$861.064 &  $-$0.206 &  28.220 &  $-$65.612   \\
TiII 4571.98 & 0.000 &  --- &  --- &  0.351 &  51.092 &  $-$53.199 &  $-$0.262 &  64.476 &  59.103   \\
TiII 5129.15 & 0.072 &  1.563 &  $-$39.754 &  0.468 &  46.196 &  $-$91.605 &  $-$0.106 &  $-$37.947 &  53.413   \\
FeI 4602.94 & 0.080 &  9.406 &  $-$12.366 &  18.393 &  110.930 &  7.339 &  $-$17.951 &  111.797 &  7.601   \\
FeII 4923.92 & 0.378 &  87.841 &  $-$38.863 &  0.137 &  54.732 &  $-$213.649 &  $-$0.331 &  54.734 &  29.902   \\
FeII 5018.43 & 0.000 &  17.804 &  $-$6.575 &  0.403 &  123.182 &  $-$28.551 &  $-$0.515 &  57.167 &  36.409   \\
MgI 5172.68 & 0.013 &  $-$0.002 &  $-$10.446 &  0.383 &  71.443 &  $-$34.529 &  $-$0.110 &  224.016 &  329.398  \\
FeI 5506.78 & 0.456 &  48.536 &  15.162 &  0.000 &  177.736 &  97.906 &  $-$0.038 &  211.393 &  217.200   \\
FeI 6200.31 & 0.522 &  66.562 &  8.013 &  0.000 &  $-$256.598 &  $-$52.042 &  $-$0.007 &  24.448 &  134.989   \\
FeI 6336.82 & 0.033 &  13.810 &  2.084 &  0.512 &  58.926 &  $-$32.655 &  $-$0.000 &  100.650 &  42.083   \\
FeI 6678.88 & 0.070 &  15.369 &  $-$0.642 &  0.636 &  73.968 &  $-$18.642 &  $-$0.000 &  80.323 &  53.022   \\
CaII 8498.02 & 0.460 &  25.756 &  $-$34.982 &  1.523 &  99.546 &  19.669 &  $-$0.206 &  12.585 &  51.413   \\
FeI 8824.22 & 0.914 &  $-$58.843 &  25.573 &  23.930 &  239.010 &  $-$63.290 &  $-$24.039 &  241.013 &  $-$62.706   \\
{\bf 2014$-$11$-$29}\\
FeI 4375.93 & 0.650 &  93.179 &  5.416 &  0.131 &  79.521 &  $-$173.971 &  $-$0.220 &  176.342 &  9.996   \\
FeI 4461.65 & 0.271 &  57.003 &  $-$25.508 &  0.121 &  230.204 &  21.772 &  $-$0.243 &  71.499 &  175.568   \\
TiII 4571.98 & 0.000 &  --- &  --- &  0.305 &  62.550 &  $-$67.851 &  $-$0.295 &  75.174 &  42.805   \\
TiII 5129.15 & 0.200 &  $-$44.092 &  $-$75.879 &  0.218 &  31.824 &  $-$128.988 &  $-$0.080 &  $-$38.721 &  33.364   \\
FeI 4602.94 & 0.328 &  $-$69.386 &  $-$21.874 &  0.115 &  33.610 &  81.528 &  --- &  --- &  ---   \\
FeII 4923.92 & 0.000 & --- &  --- &  0.374 &  138.202 &  $-$48.961 &  $-$0.333 &  76.867 &  48.553   \\
FeII 5018.43 & 0.000 &  --- &  --- &  0.425 &  89.905 &  $-$78.094 &  $-$0.242 &  155.300 &  15.616   \\
FeI 5332.90 & 0.000 &  --- &  --- &  0.226 &  98.507 &  33.966 &  $-$0.220 &  143.016 &  124.855   \\
FeII 5991.38 & 0.049 &  1.713 &  $-$17.621 &  8.185 &  96.472 &  72.551 &  $-$8.047 &  97.475 &  73.867   \\
FeI 6200.31 & 0.000 &  --- &  --- &  0.267 &  70.920 &  18.415 &  $-$0.059 &  14.277 &  146.861   \\
FeI 6336.82 & 0.402 &  94.743 &  $-$29.822 &  0.296 &  218.282 &  318.075 &  $-$0.304 &  86.382 &  260.118   \\
SiII 6347.10 & 0.082 &  0.061 &  $-$10.631 &  0.343 &  65.176 &  $-$101.036 &  $-$0.191 &  183.097 &  $-$33.567   \\
FeI 6393.60 & 0.215 &  35.831 &  53.055 &  0.328 &  $-$45.185 &  $-$12.247 &  $-$0.122 &  32.709 &  174.774   \\
FeI 6678.88 & 0.100 &  $-$29.709 &  29.222 &  12.371 &  83.582 &  23.566 &  $-$12.119 &  82.705 &  24.560   \\
CaII 8498.02 & 0.042 &  0.072 &  10.793 &  2.938 &  103.286 &  29.733 &  $-$1.883 &  81.478 &  44.669   \\
FeI 8824.22 & 0.598 &  58.689 &  21.960 &  74.273 &  255.921 &  $-$33.838 &  $-$74.317 &  256.667 &  $-$33.798   \\
{\bf 2014$-$12$-$04}\\
FeI 4461.65 & 0.382 &  42.051 &  $-$22.172 &  0.206 &  39.489 &  $-$84.491 &  $-$0.180 &  $-$29.410 &  213.730   \\
FeI 4375.93 & 0.438 &  76.279 &  5.125 &  0.125 &  67.243 &  $-$189.413 &  $-$0.182 &  34.239 &  40.331   \\
TiII 4571.98 & 0.217 &  3.200 &  $-$8.009 &  0.290 &  56.251 &  $-$78.774 &  $-$0.319 &  59.028 &  32.945   \\
FeII 4923.92 & 0.312 &  46.476 &  $-$41.916 &  0.184 &  38.318 &  $-$108.037 &  $-$0.274 &  71.278 &  38.468   \\
FeII 5018.43 & 0.328 &  58.150 &  $-$60.574 &  0.190 &  31.187 &  $-$137.174 &  $-$0.253 &  70.381 &  27.730   \\
TiII 5129.15 & 3.532 &  26.895 &  42.631 &  0.251 &  38.023 &  $-$112.646 &  $-$3.610 &  27.617 &  42.527   \\
MgI 5167.32 & 0.013 &  $-$0.195 &  $-$11.062 &  0.584 &  90.115 &  $-$35.108 &  $-$0.169 &  $-$30.784 &  137.500   \\
MgI 5172.68 & 0.201 &  57.040 &  $-$58.845 &  0.000 &  $-$38.962 &  $-$71.687 &  $-$0.009 &  14.302 &  50.641   \\
FeI 5332.90 & 0.150 &  77.937 &  $-$20.348 &  0.106 &  14.743 &  77.679 &  $-$0.106 &  126.035 &  132.409   \\
FeI 5506.78 & 0.081 &  48.628 &  $-$4.177 &  --- &  --- & --- &  $-$0.073 &  84.799 &  169.557   \\
FeII 5991.38 & 0.037 &  20.464 &  $-$1.976 &  0.322 &  148.142 &  182.913 &  $-$0.401 &  112.559 &  183.544   \\
FeI 6191.56 & 90.659 &  212.369 &  $-$61.654 &  0.479 &  141.217 &  $-$236.672 &  $-$90.672 &  213.350 &  $-$62.270   \\
FeI 6200.31 & 0.183 &  64.188 &  $-$5.453 &  0.413 &  23.683 &  93.229 &  $-$0.382 &  22.070 &  95.660   \\
FeI 6336.82 & 0.000 &  --- &  --- &  0.373 &  118.427 &  2.018 &  $-$0.191 &  74.471 &  36.868   \\
SiII 6347.10 & 0.000 & --- &  --- &  0.148 &  58.763 &  $-$110.175 &  $-$0.086 &  115.528 &  16.038   \\
FeI 6400.00 & 0.799 &  67.141 &  2.567 &  9.974 &  100.290 &  82.612 &  $-$10.085 &  103.054 &  80.635   \\
FeI 6678.88 & 0.227 &  65.095 &  $-$44.398 &  0.141 &  75.219 &  80.238 &  $-$0.066 &  $-$33.271 &  85.017   \\
\hline
\end{longtable}

\section{Other correlations between line velocity and atomic parameters \label{correlations-appendix}}

\begin{figure*}
\centering
\includegraphics[width=15cm]{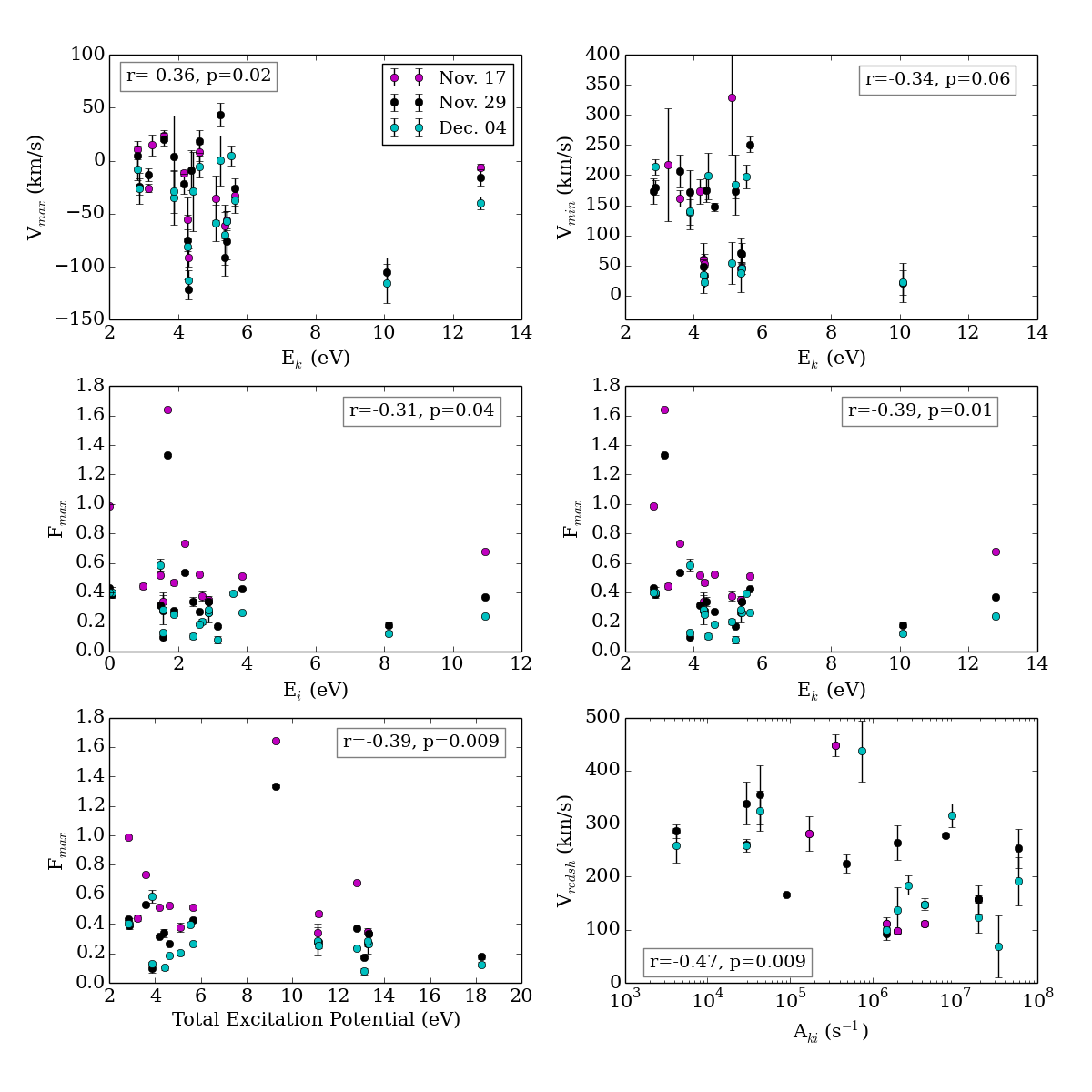}
\caption{Marginal correlations observed between the velocity data and different atomic parameters. See text for
details.}
\label{mcorrelation-fig}%
\end{figure*}

\begin{figure*}
\centering
\includegraphics[width=15cm]{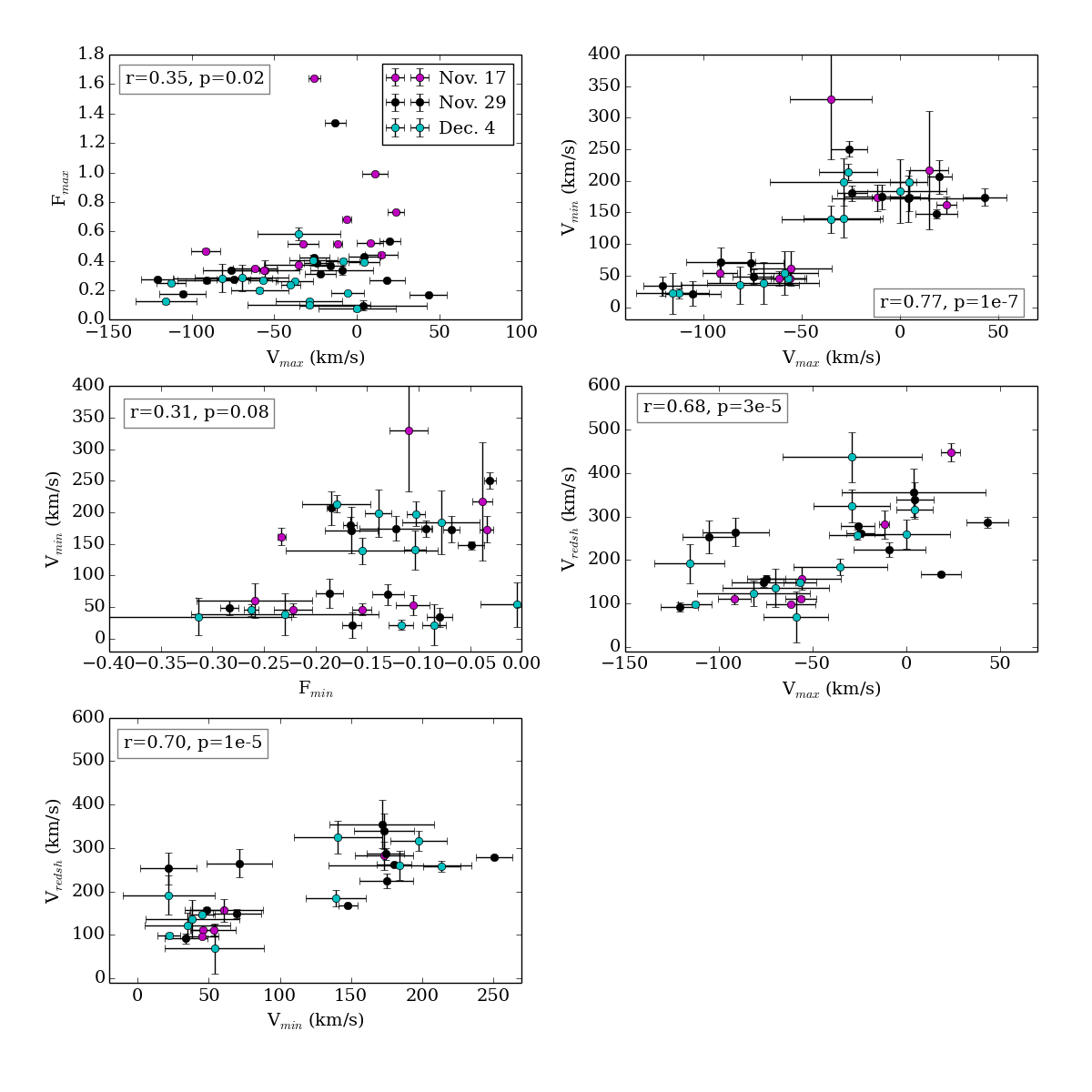}
\caption{Correlations between the different line parameters. We note that the correlations between the peak, minimum, and
maximum red-shifted velocities simply indicate that the lines are usually globally shifted.}
\label{lcorrelation-fig}%
\end{figure*}

Figure \ref{mcorrelation-fig} displays the marginal correlations observed
between the line parameters and the upper and lower energy of the levels (E$_k$ and E$_i$) and the
sum of the ionization and excitation potential. Figure \ref{lcorrelation-fig} shows the correlations between the line parameters
themselves, some of which are trivial, such as the correlation between all the different line velocities, which
indicate that the lines tend to be globally shifted.

\end{appendix}

\end{document}